\documentclass[reprint,aps,prl,
groupedaddress,superscriptaddress
]{revtex4-2}
\usepackage{graphicx}
\usepackage{xcolor}
\definecolor{codexdarkgreen}{RGB}{0,100,0}
\definecolor{codexdeepviolet}{RGB}{102,0,153}
\definecolor{codexteal}{RGB}{0,96,112}
\definecolor{codexsienna}{RGB}{139,69,19}
\definecolor{codexburgundy}{RGB}{128,0,64}
\definecolor{codexorange}{RGB}{220,90,0}
\definecolor{codexcobalt}{RGB}{0,70,160}
\definecolor{codexrose}{RGB}{190,60,110}
\definecolor{codexmagenta}{RGB}{155,0,115}
\definecolor{codexindigo}{RGB}{55,55,145}
\definecolor{codexforest}{RGB}{0,105,85}
\definecolor{codexochre}{RGB}{140,100,0}
\usepackage{dcolumn}
\usepackage{rotating}
\usepackage{makecell}
\usepackage{enumerate} 
\usepackage{bm}
\usepackage{bbm}
\usepackage{amsmath}
\usepackage{amsfonts}
\usepackage{tikz}
\usepackage[version=4]{mhchem}
\usepackage{adjustbox}
\usepackage[justification=Justified]{caption}
\usepackage[justification=Justified]{subcaption}
\usepackage{ragged2e}
\usepackage{booktabs}
\usepackage{placeins}
\usepackage{siunitx}
\usepackage{multirow}
\usepackage{tabularx}
\usepackage{tikz}
\usepackage{circuitikz}
\usetikzlibrary{patterns}
\usetikzlibrary{decorations.markings,  decorations.pathmorphing, decorations.pathreplacing, decorations.shapes, arrows.meta}

\bibpunct{}{}{,}{s}{}{\textsuperscript{,}}
\newcommand{\Vdot}[1]{\draw[fill] #1 circle[radius=4pt];}

\newcommand{\Vline}[3]{\draw[shift={#1},rotate={#2},decorate,decoration={coil,aspect=0}] (0,0) -- (#3,0) ;}
\newcommand{\Wline}[3]{\draw[shift={#1},rotate={#2},decorate,double,decoration={coil,aspect=0}] (0,0) -- (#3,0) ;}
\newcommand{\cmdot}[2][6]{\draw[shift={#2}, fill={rgb, 255:red, 208; green, 2; blue, 27}, fill opacity=1] (-#1pt,-#1pt) rectangle (#1pt,#1pt);}

\newcommand{\gfarrow}{ \arrow[thick,scale=-1.2]{Stealth[bend]} }
\newcommand{\gfarrowpos}{.43}

\newcommand{\gfline}[4][50]{\draw[shift={#2},rotate={#3},thick,postaction={decorate},decoration={markings,mark=at position \gfarrowpos with { \gfarrow }}] (0,0) to[out=-#1,in=#1-180] (#4,0);}
\newcommand{\gflinex}[4][50]{\draw[shift={#2},rotate={#3},color=red,thick,postaction={decorate},decoration={markings,mark=at position \gfarrowpos with { \gfarrow }}] (0,0) to[out=-#1,in=#1-180] (#4,0);}

\newcommand{\phbbl}[4][50]{\gfline[#1]{#2}{#3}{#4};\draw[shift={#2},rotate={#3},thick,postaction={decorate},decoration={markings, mark=at position \gfarrowpos with { \gfarrow }}] (#4,0) to[out=180-#1,in=#1] (0,0);}
\newcommand{\phbblx}[4][50]{\gflinex[#1]{#2}{#3}{#4};\draw[shift={#2},rotate={#3},color=red,thick,postaction={decorate},decoration={markings, mark=at position \gfarrowpos with { \gfarrow }}] (#4,0) to[out=180-#1,in=#1] (0,0);}

\newcommand{\phbblc}[4][50]{\phbbl[#1]{#2}{#3}{#4} 
\pattern [shift={#2},rotate={#3},pattern=north east lines] (0,0) to[out=-#1,in=#1-180] (#4,0) to[in=#1,out=180-#1] cycle; }
\newcommand{\phbblcx}[4][50]{\phbblx[#1]{#2}{#3}{#4} 
\pattern [shift={#2},rotate={#3},pattern color=red,pattern=north east lines] (0,0) to[out=-#1,in=#1-180] (#4,0) to[in=#1,out=180-#1] cycle; }

\newcommand{\gfloop}[3]{\draw[shift={#1},rotate={#2},thick,postaction={decorate},decoration={markings,mark=at position 0.05 with { \gfarrow }}] (#3,0) circle[radius=#3];}

\newcommand{\ii}{\mathbbm{i}}
\newcommand{\hH}{\hat{H}}

\newcommand{\hp}{\hat{p}}
\newcommand{\hq}{\hat{q}}
\newcommand{\hr}{\hat{r}}
\newcommand{\hs}{\hat{s}}
\newcommand{\hx}{\hat{x}}
\newcommand{\hy}{\hat{y}}
\newcommand{\hz}{\hat{z}}

\newcommand{\eriph}[2]{\langle#1|#2\rangle}
\newcommand{\erias}[2]{\langle#1||#2\rangle}

\newcommand{\tord}{\mathcal{T}}
\newcommand{\Tau}[1]{\tord[#1]}

\newcommand{\spinarrow}{ \arrow[thick,scale=1.]{Stealth[bend]} }
\newcommand{\spinarrowpos}{1.}
\newcommand{\spinup}[2]{\draw[shift={#1},thick,postaction={decorate},decoration={markings,mark=at position \spinarrowpos with { \spinarrow }}] (0,0) to (0,#2);}
\newcommand{\spindown}[2]{\draw[shift={#1},thick,postaction={decorate},decoration={markings,mark=at position \spinarrowpos with { \spinarrow }}] (0,#2) to (0,0);}

\renewcommand{\Re}{\mathrm{Re}}

\newcommand{\mycolor}{black}

\begin{document}


\title{
{\color{\mycolor}A Diagrammatic Route to Multi-Reference $GW$ for Strongly Correlated Molecules}
}
\author{Yuqi Wang}
\author{Wei-Hai Fang}
\author{Zhendong Li}\email{zhendongli@bnu.edu.cn}
\affiliation{Key Laboratory of Theoretical and Computational Photochemistry, Ministry of Education, College of Chemistry, Beijing Normal University, Beijing, 100875, China}
\affiliation{Institute for Advanced Study, Beijing Normal University, Beijing, 100875, China}

\begin{abstract}
The $GW$ approximation is a cornerstone of many-body perturbation theory for computing single-particle excitations, yet it can fail qualitatively in strongly correlated systems where the single-reference picture becomes inadequate.
{\color{\mycolor}
To overcome this limitation, we develop a diagrammatic route to multi-reference $GW$ (MR-$GW$) from an interacting active-space reference, treating strong correlation nonperturbatively and the residual interaction through dynamical screening at the multi-reference random phase approximation level. 
Because the conventional diagrammatic self-energy expansion no longer applies, we define a practical one-shot MR-$GW$ approximation within a generalized diagrammatic framework, recovering standard $GW$ as a special case.} Applications to prototypical correlated atoms and molecules show that the interacting active-space reference captures multiconfigurational satellite structures poorly described by single-reference $GW$, while dynamical screening generally improves the satellite energies as well as the leading ionization and electron-attachment energies. This work opens a diagrammatic route to Green's-function methods that combine nonperturbative strong correlation with dynamical screening.
\end{abstract}


\maketitle

\textit{Introduction---}Green's functions provide a direct description of electron addition and removal, connecting many-body theory to photoelectron spectra.\cite{onida_electronic_2002} Most practical approaches are
developed under the framework of many-body perturbation theory (MBPT).\cite{martin_interacting_2016,fetter_quantum_1971} 
The standard paradigm is to approximate the self-energy $\Sigma$,
which by the Dyson equation\cite{dyson_s_1949} relates the interacting Green's function $G$ to a non-interacting zeroth-order Green's function $G_0$ defined by a \emph{quadratic} zeroth-order Hamiltonian $\hH_0$.
The $GW$ approximation is a central success of this strategy: 
it combines a one-particle propagator with a dynamically screened Coulomb interaction $W$ and has become a standard tool for charged excitations in molecules and materials.\cite{hedin_new_1965,hybertsen_electron_1986,aryasetiawan_gw_1998,hedin_correlation_1999,leng_gw_2016,reining_gw_2018,golze_gw_2019,ren_resolution--identity_2012,jiang_fhi-gap_2013,huser_quasiparticle_2013,gulans_exciting_2014,van_setten_gw100_2015,govoni_large_2015,bruneval_molgw_2016,bruneval_gw_2021,zhu_all-electron_2021,zhang_many-body_2023} 
{\color{\mycolor}Throughout this Letter, $GW$ denotes the one-shot $G_0W_0$
approximation unless stated otherwise, in which the self-energy is evaluated from a fixed $G_0$ and its corresponding screened interaction $W_0$.}

Despite its widespread success, $GW$ has well-documented theoretical and practical limitations.\cite{romaniello_self-energy_2009, caruso_bond_2013,reining_gw_2018,ammar_can_2024}
One of the most significant challenges is its failure in the presence of strong correlation, which is difficult to capture perturbatively.
Among other manifestations of strong correlation\cite{mott_basis_1949,hubbard_electron_1963,loos_uniform_2016,wigner_interaction_1934}, 
multi-configurational molecules represent a typical example, 
where the ground-state wavefunction contains several configurations with comparable weights. They are encountered in a wide range of important systems
including stretched bonds, radicals, conjugated systems, transition-metal complexes, and defects.\cite{lyakh_multireference_2012,cederbaum_many-body_1980,mejuto-zaera_are_2021}
The performance of $GW$ in multi-configurational systems has recently been tested systematically, revealing a series of qualitative errors.\cite{ammar_can_2024}

While previous attempts to address strong correlation 
have largely relied on integrating $GW$ into quantum
embedding frameworks\cite{lee2017diatomic,lan2017testing,sheng_greens_2022},
we propose a fundamentally different framework rooted in a quantum chemistry perspective. The lesson well learned in quantum chemistry is that multi-configurational systems can be treated by multi-reference perturbation theories (MRPT)\cite{andersson_secondorder_1992,angeli_introduction_2001,park_multireference_2020}, {\color{\mycolor}which employ an interacting $\hH_0$ to capture  correlation effects among a set of strongly correlated orbitals, referred to as active space, at the zeroth order, and treat the residual interaction perturbatively.}
Transferring this principle to Green's-function MBPT is highly non-trivial\cite{brouder_structure_2009}, 
because Wick's theorem,\cite{wick_evaluation_1950} which relies on quadratic $\hat{H}_0$ and a determinantal reference, breaks down for interacting $\hat{H}_0$.
Consequently, a series of fundamental tools in MBPT, most notably the Dyson equation and Hedin's equations, cannot be carried over unchanged.

{\color{\mycolor}
In this Letter, we make two advances in this direction. First, we develop a systematic diagrammatic multi-reference MBPT framework, which includes strong correlation at the zeroth order nonperturbatively via an interacting $\hat{H}_0$, and treats the residual weak interaction perturbatively by combining diagrammatic expansion\cite{metzner1991linked,wang_generalized_2025,wang_unified_2025}
with the generalized Dyson equation.\cite{brouder_structure_2009,hall_non-equilibrium_1975}
Second, we introduce a practical one-shot $GW$-like approximation within this framework using the same $GW$ self-energy diagram, referred to as the MR-$GW$ approximation.
This construction combines nonperturbative strong correlation with dynamical screening and recovers standard $GW$ exactly as a special case.}
Numerical results show that MR-$GW$ improves spectral functions, recovers satellites missed by single-reference $GW$, and restores the correct ordering of ionized states
in representative strongly correlated atoms and molecules.
This work establishes a new theoretical platform for developing ab-initio Green's function methods for strongly correlated systems.
 
\textit{Multi-reference theoretical framework---}{\color{\mycolor}
For strongly correlated systems, an appropriate zeroth-order Hamiltonian $\hH_0$ explicitly retains the two-electron interactions within the active space that are responsible for strong correlation.\cite{sokolov_chapter_2024}} We choose the Dyall Hamiltonian\cite{dyall_choice_1995}, widely used in MRPT as $\hat{H}_0$, and partition the total set of spin-orbitals (labeled by $\{p,q,r,s,\cdots\}$) into an active subset (labeled by $\{w,x,y,z,\cdots\}$) that carries the strong correlation and an inactive subset (labeled by $\{P,Q,R,S,\cdots\}$) comprising the remaining doubly occupied core orbitals and unoccupied virtual orbitals (Fig.~\ref{fig:framework}a):
\begin{align}
    \hH_{0}^\text{Dyall} &= \epsilon_P \hat{P}^\dagger \hat{P}
    + \hH_0^{\text{act}}, \nonumber\\
    \hH_0^{\text{act}} &=
    h^{\text{eff}}_{xy}\hat{x}^{\dagger}\hat{y} + \hat{V}^{A}, \nonumber\\
    \hat{V}^{A} &= \frac{1}{2}
    \eriph{xy}{zw} \hat{x}^{\dagger}\hat{y}^{\dagger}\hat{w}\hat{z}\label{eq:Hdyall-main2}
\end{align}
Here $\hat{p}^{(\dagger)}$ annihilates (creates) a fermion and repeated indices are summed. The full Coulomb interaction $\hat V^A$ is retained among active orbitals, while the inactive part is quadratic in $\hH_{0}^\text{Dyall}$. 
The residual interaction is assumed to be relatively weak and is treated perturbatively:
\begin{eqnarray}
\hat{V} = \hat{H} - \hH_{0}^{\text{Dyall}} = 
u_{pq}\hat{p}^\dagger \hat{q}
+
\frac{1}{2}v_{pr,qs}\hat{p}^\dagger\hat{q}^\dagger\hat{s}\hat{r}.\label{eq:PT-V}
\end{eqnarray}
Detailed expressions of $\epsilon_P$, $h_{xy}^\text{eff}$, $u_{pq}$ and $v_{pr,qs}$ are given in Appendix A.

The additive structure of $\hat{H}^{\mathrm{Dyall}}_0$ factorizes each zeroth-order eigenstate as $|\Phi_\mu\rangle=|\Phi_{\mu_I}^I\rangle|\Xi_{\mu_A}^A\rangle$. The inactive part $|\Phi_{\mu_I}^I\rangle$ is a single determinant as in standard MBPT,  {\color{\mycolor}
whereas the active part $|\Xi_{\mu_A}^A\rangle$ is a general multiconfigurational state $|\Xi_{\mu_A}^A\rangle=\sum_\lambda|\Phi_\lambda^A\rangle C_{\lambda,\mu_A}$ in the occupation-number (Slater-determinant) basis
$\{|\Phi_\lambda^A\rangle\}$ of the active space.}
Consequently, the zeroth-order one-body Green's function is block diagonal in orbital space
\begin{align}
    \mathbf{G}_0(\omega) = \begin{bmatrix}
        \mathbf{G}_0^I(\omega) & \mathbf{0} \\
        \mathbf{0}     & \mathbf{G}_0^A(\omega)
    \end{bmatrix},\label{eq:MRGW-G0}
\end{align}
where bold symbols denote matrices. The inactive block $\mathbf{G}_{0}^{I}$ is simply a non-interacting Green's function
\begin{align}
[\mathbf{G}^I_0]_{PQ}(\omega) &= (\omega - \epsilon_P + \text{sgn}(P)\ii 0^+)^{-1}\delta_{PQ},
\end{align}
with $\text{sgn}(P)$ being $-1$ ($+1$) for (un)occupied orbitals. In contrast, the active part $\mathbf{G}_0^A$ is interacting and contains the many-body addition and removal poles generated inside the active space.

\begin{figure}[t]
    \centering

    \begin{subfigure}{.3\linewidth}
    \caption{\hspace{\linewidth} }
    \vspace{-10pt}

    \centering
    
    \begin{tikzpicture}[scale=.435, trim right=2cm, trim left=-.8cm]

\draw[very thick] (0,0) -- (3,0);
\draw[very thick] (0,1.5) -- (3,1.5);

\draw[very thick, color=red] (0,3) -- (3,3);
\draw[very thick, color=red] (0,4.5) -- (3,4.5);

\draw[very thick] (0,6) -- (3,6);
\draw[very thick] (0,7.5) -- (3,7.5);

\spinup{(1,0)}{1.2}
\spindown{(2,0)}{1.2}

\spinup{(1,1.5)}{1.2}
\spindown{(2,1.5)}{1.2}

\draw[very thick] (1.5,3+.3) circle[radius=.3];
\draw[very thick] (1.5,4.5+.3) circle[radius=.3];

\end{tikzpicture}
    \end{subfigure}
    \hfill
    \begin{subfigure}{.68\linewidth}

    \begin{subfigure}{\linewidth}
    \caption{\hspace{\linewidth} }

    \vspace{-8pt}
    
    
    \begin{tikzpicture}
        \node at (-1,0) {$\Sigma^{\text{xc},GW} =$};
        \gfline[-45]{(0,0)}{0}{2}
        \Wline{(0,0)}{0}{2}
        \node at (1.5,.55) {$G_0$};
        \node at (1.5,-.25) {$W$};
    \end{tikzpicture}
    \end{subfigure}
    \vfil
    \begin{subfigure}{\linewidth}
    \caption{\hspace{\linewidth} }
    \vspace{-16pt}

    \flushright
    
    \begin{tikzpicture}[trim left=0cm]
        \Wline{(0,0)}{-90}{1}
        \node at (1,-.4) {$=$};
        \node at (.3,-.8) {$W$};
        \Vline{(2,0)}{-90}{1}
        \node at (2.2,-.8) {$v$};
        \node at (3,-.4) {$+$};
        \Vline{(4,.6)}{-90}{.6}
        \phbblc{(4,0)}{-90}{.6}
        \node at (4.5,-.3) {$\Pi_0$};
        \Wline{(4,-.6)}{-90}{.6}
    \end{tikzpicture}
    \end{subfigure}

    \end{subfigure}

   \begin{subfigure}{\linewidth}
   \flushright
    \caption{\hspace{\linewidth} }
    \vspace{-4pt}
    \begin{tikzpicture}[scale=.34]
        \phbblc{(0.,0)}{-90}{3.4}
        \node at (2.5,-1.7) {$=$};
        \phbbl{(5,0)}{-90}{3.4}
        \node at (7.5,-1.7) {$+$};
        \gflinex{(10,0)}{-90}{3.4}
        \gfline{(10,-3.4)}{90}{3.4}
        \node at (12.5,-1.7) {$+$};
        \gfline{(15,0)}{-90}{3.4}
        \gflinex{(15,-3.4)}{90}{3.4}
        \node at (17.5,-1.7) {$+$};
        \phbblcx{(20,0)}{-90}{3.4}

        \node at (1.1,0) {$\Pi_0$};

        \node at (6.5,-.5) {$G_0^I$};
        \node at (3.5,-2.9) {$G_0^I$};

        \node at (6.5+5,-.5) {$G_0^I$};
        \node at (3.4+5,-2.9) {$G_0^A$};

        \node at (6.6+10,-.5) {$G_0^A$};
        \node at (3.5+10,-2.9) {$G_0^I$};

        \node at (21.7,-.5) {$\Pi_0^A$};
        
    \end{tikzpicture}
    \end{subfigure}

    \caption{
    Schematic representation of the MR-$GW$ approach.
    (a) Spin-orbitals are divided into active orbitals (red)
    and inactive (black) orbitals in the multi-reference framework.
    (b) $GW$ self-energy diagram: directed line represents $G_0$, while the double wiggly line represents the screened interaction $W$.
    (c) Screened interaction: wiggly lines represent either the full Coulomb interaction in the standard $GW$ or the residual two-electron interaction \eqref{eq:PT-V} in MR-$GW$.
    (d) Irreducible polarizability in MR-RPA\cite{wang_generalized_2025} (active-space quantities in red): 
    the first three terms are products of $G_0$, familiar from the standard RPA, while the last term represents the active-space irreducible polarizability. 
    }
\label{fig:framework}
\end{figure}

\textit{Generalized Dyson equation---}In standard MBPT, the Dyson equation
connects the exact Green's function $\mathbf{G}$ to the zeroth-order $\mathbf{G}_0$ through a single self-energy $\mathbf{\Sigma}$. However, when $\hH_0$ is interacting or the zeroth-order state is multi-configurational,
the conventional diagrammatic expansion based on
Wick's theorem and the Dyson equation
does not apply,\cite{brouder_structure_2009} 
and must be supplemented by connected many-body reference Green’s functions\cite{negele_quantum_1998,metzner1991linked}. A useful organization of the resulting expansion is provided by the generalized Dyson structure,\cite{brouder_structure_2009}
\begin{align}
    \mathbf{M} &= (\mathbf{I}-\mathbf{\Sigma}^{21})^{-1} (\mathbf{G}_0+\mathbf{\Sigma}^{22})(\mathbf{I}-\mathbf{\Sigma}^{12})^{-1},
    \nonumber\\
    \mathbf{G} &= \mathbf{M}  + \mathbf{M}\mathbf{\Sigma}^{11}\mathbf{G},\label{eq:GDE-11}
\end{align}
originally derived by Hall for systems with initial correlation\cite{hall_non-equilibrium_1975} and later by Brouder via an auxiliary-source construction\cite{brouder_structure_2009}. The frequency argument $\omega$ in Eq. \eqref{eq:GDE-11} is suppressed, and the four matrices $\mathbf{\Sigma}^{ij}$ are one-particle irreducible (1PI) generalizations of the self-energy. This complexity leaves the structure of Green's functions built on an interacting reference far less explored than that of their single-reference counterparts. To the best of our knowledge, diagrams for $\mathbf{\Sigma}^{ij}$ have only been derived for simple interacting quantum field theories\cite{brouder_structure_2009}.

We present the first-order diagrammatic expansions of $\mathbf{\Sigma}^{ij}$ in Appendix B for the perturbation defined in Eq. \eqref{eq:PT-V}, which suffice to establish the MR-$GW$ approximation below. Such an expansion can be carried out to higher orders, and the general diagrammatic structures are as follows:
For $\mathbf{\Sigma}^{11}$ \eqref{eq:GDE-11}, 
besides the conventional diagrams built from $\mathbf{G}_0$,
there are additional generalized Feynman diagrams involving also
connected many-body Green's functions\cite{negele_quantum_1998,metzner1991linked} (cumulants) in the active space, whereas the other three self-energies $\mathbf{\Sigma}^{12}$, $\mathbf{\Sigma}^{21}$, and $\mathbf{\Sigma}^{22}$ 
consist entirely of such generalized diagrams. 
For an empty active space or $\hat V^A=0$ with a Slater-determinant
reference, zeroth-order cumulants and the three
additional $\mathbf{\Sigma}^{ij}$ vanish, 
recovering the conventional Dyson equation.

The generalized Dyson equation \eqref{eq:GDE-11}, along with the diagrammatic expansion for self-energies, forms a new theoretical platform for developing practical 
Green's function methods by approximating $\mathbf{\Sigma}^{ij}$. {\color{\mycolor}This framework provides a diagrammatic route to a multireference
generalization of $GW$ through screened-interaction resummation,
without requiring a direct extension of Hedin's equations to the
interacting reference.}

\textit{Multi-reference $GW$ approximation---} Diagrammatically, the standard $GW$ approximates the self-energy by replacing the bare Coulomb interaction in the first-order exchange diagram with the screened interaction $W$, giving $\Sigma^{\text{xc},GW}=\ii GW$ [Fig.~\ref{fig:framework}(b)] for the exchange-correlation self-energy\cite{hedin_new_1965}.
The screened interaction [Fig.~\ref{fig:framework}(c)] is introduced by\cite{hubbard1957description}
\begin{align}
    \mathbf{W}(\omega) = \mathbf{v} + \mathbf{v}\mathbf{\Pi}(\omega)\mathbf{v} 
    = \mathbf{v} + \mathbf{v}\mathbf{\Pi}_0(\omega)\mathbf{W}(\omega),\label{eq:dyson-W}
\end{align}
where $v_{pr,qs}=\langle pq|rs\rangle$ represents
the full Coulomb interaction, and the reducible polarizability $\mathbf{\Pi}(\omega)$ is evaluated at the random phase approximation\cite{pines_collective_1952,bohm_collective_1953,gell-mann_correlation_1957} (RPA) level
\begin{eqnarray}
    \mathbf{\Pi}(\omega) = \mathbf{\Pi}_0(\omega) + \mathbf{\Pi}_0(\omega)\mathbf{v}\mathbf{\Pi}(\omega),\label{eq:Pi}
\end{eqnarray}
with $\mathbf{\Pi}_0(\omega)$ the zeroth-order irreducible polarizability. 
{\color{\mycolor}
Using the above diagrammatic framework, we naturally generalize one-shot $GW$ as follows:

First, the screened interaction $W$ is evaluated
at the multi-reference RPA (MR-RPA) level\cite{wang_generalized_2025},
which accounts for the four kinds of 
screening effects shown in Fig. \ref{fig:framework}(d).}
This amounts to constructing $W$ using Eq. \eqref{eq:dyson-W}, with the full Coulomb interaction replaced by the residual two-electron interaction in Eq. \eqref{eq:PT-V} and the noninteracting $\mathbf{\Pi}_0(\omega)$ of standard RPA replaced by the corresponding quantity for the interacting $\hH_0^{\mathrm{Dyall}}$,
\begin{align}
[\mathbf{\Pi}_0]_{pr,qs}(\omega)  \equiv &\sum_{\mu>0} \frac{\langle \Phi_0|\hp^\dagger \hr|\Phi_\mu\rangle \langle \Phi_\mu|\hq^\dagger \hs|\Phi_0\rangle}{\omega - \omega_{\mu} + \ii0^+} 
\nonumber\\
&- \sum_{\mu>0}
\frac{\langle \Phi_0|\hq^\dagger \hs|\Phi_\mu\rangle \langle \Phi_\mu|\hp^\dagger \hr|\Phi_0\rangle }
{\omega + \omega_{\mu} - \ii0^+}, \label{eq:mr-rpa-pi0}
\end{align}
where $\omega_\mu$ is the excitation energy of the zeroth-order excited state $|\Phi_{\mu}\rangle$.

{\color{\mycolor}
Second, we define the practical MR-$GW$ approximation by retaining,
within the $\mathbf{\Sigma}^{11}$ block, the same self-energy diagrams that enter
standard $GW$. Specifically, $\mathbf{\Sigma}^{11}$ contains the
first-order one-electron perturbation and Hartree contributions
[Fig.~\ref{fig:GDE-1}(a) in Appendix B], together with the standard
$GW$ self-energy diagram [Fig.~\ref{fig:framework}(b)]. For
the Dyall Hamiltonian, the first-order contribution
$\mathbf{\Sigma}^{22}_1$ vanishes. The present approximation neglects
the mixed blocks $\mathbf{\Sigma}^{12/21}$, together with
higher-order contributions to $\mathbf{\Sigma}^{22}$.
We therefore denote this selected approximation to
$\mathbf{\Sigma}^{11}$ by
$\mathbf{\Sigma}^{\text{MR-}GW}$, whose algebraic form is derived in
Eq.~\eqref{eq:sigma-mrgw-general} of the Supplemental
Material.\cite{SM}
}
Under this block structure, the generalized Dyson equation,
Eq.~\eqref{eq:GDE-11}, reduces to
\begin{align}
    \mathbf{G}(\omega)
    =
    \mathbf{G}_0(\omega)
    +
    \mathbf{G}_0(\omega)
    \mathbf{\Sigma}^{\text{MR-}GW}(\omega)
    \mathbf{G}(\omega),
    \label{eq:dyson-mrgw}
\end{align}
from which the spectral function $A(\omega)$ is obtained.

{\color{\mycolor}
The resulting MR-$GW$ approximation is a minimal and
natural multireference extension of one-shot $GW$. 
By retaining the
screened-interaction self-energy structure of standard $GW$, 
the frequency-dependent part of
$\mathbf{\Sigma}^{\text{MR-}GW}(\omega)$ admits a simple-pole
representation with positive-semidefinite residue matrices
for a stable MR-RPA solution, consistent
with the Lehmann representation. Consequently, the resulting Green's
function preserves the appropriate causal analytic structure and
yields a nonnegative spectral function $A(\omega)$. 
The consequences of retaining the mixed blocks
$\mathbf{\Sigma}^{12/21}$ 
are examined in Appendix C.}

\textit{Connection to standard self-energy---}To better understand the above MR-$GW$ approximation, we can recast it as an approximation to the standard self-energy. Introduce a noninteracting Green's function $\mathbf{g}_0$ and define $\mathbf{\Sigma}_{0}$ so that it connects $\mathbf{g}_0$ to the interacting reference $\mathbf{G}_0$ of Eq.~\eqref{eq:MRGW-G0} via the standard Dyson equation:
\begin{eqnarray}
    \mathbf{\Sigma}_0[\mathbf{g}_0,\hat{V}^A] = \mathbf{g}_0^{-1}-\mathbf{G}_0^{-1}. \label{eq:dyson-G0}
\end{eqnarray}
Here, the bracket in $\mathbf{\Sigma}_0[\mathbf{g}_0,\hat{V}^A]$ emphasizes that $\mathbf{\Sigma}_0$ is a functional of $\mathbf{g}_0$ and $\hat{V}^A$,
which reflects the nonperturbative reference correlation due to $\hat{V}^A$ (the difference between the mean fields used to define $\mathbf{g}_0$ and $\mathbf{G}_0$ is suppressed for clarity). 
Combining Eqs. \eqref{eq:dyson-mrgw} and \eqref{eq:dyson-G0} recovers the standard Dyson equation relating $\mathbf{G}$ and $\mathbf{g}_0$,
\begin{align}
    \mathbf{G} = \mathbf{g}_0 + \mathbf{g}_0\left(
    \mathbf{\Sigma}_0[\mathbf{g}_0,\hat{V}^A] 
    + \mathbf{\Sigma}^{\text{MR-}GW}[\mathbf{G}_0,\mathbf{\Pi}_0,\hat{V}]
    \right)\mathbf{G}. \label{eq:dyson-unified}
\end{align}
{\color{\mycolor}
Since $\mathbf{G}_0$ and $\mathbf{\Pi}_0$ are also functionals of $\mathbf{g}_0$ and $\hat{V}^A$, the parenthesized terms in Eq. \eqref{eq:dyson-unified} constitute the corresponding self-energy of MR-$GW$ in the standard MBPT picture.
Eq. \eqref{eq:dyson-unified} reveals two important differences between
$GW$ and MR-$GW$: $\mathbf{\Sigma}_0$ responsible for strong correlation inside the active space is completely missing in $GW$, whereas $\mathbf{\Sigma}^{\text{MR-}GW}$ includes the screened residual interaction at the MR-RPA level rather than the full Coulomb interaction. Consequently, unlike in many $GW$ embedding schemes, no separate double-counting correction is required.\cite{lee2017diatomic,lan2017testing,sheng_greens_2022}}

\textit{Results---}MR-$GW$ is implemented\cite{MRMBPTcode} {\color{\mycolor}using molecular integrals and a zeroth-order reference obtained from
complete-active-space configuration-interaction (CASCI) or
complete-active-space self-consistent-field (CASSCF)
calculations\cite{lyakh_multireference_2012} performed with the
\textsc{PySCF} package.\cite{sun_recent_2020}
The present pilot implementation uses an explicit spectral
representation for $\mathbf{\Sigma}^{\text{MR-}GW}(\omega)$,
obtained by full diagonalization of the MR-RPA problem,\cite{wang_generalized_2025}
see the Supplemental Material\cite{SM}.
For a fixed active space, it retains the same asymptotic
orbital-space scaling as the corresponding dense spectral implementation
of standard $G_0W_0$, while introducing additional costs associated
with the interacting active-space reference.
The computational workflow and its cost relative to standard $GW$ are
summarized in Appendix D.}

We apply MR-$GW$ to a set of prototypical strongly correlated systems
and benchmark its predictions against {\color{\mycolor}
standard $GW$ calculations based on RHF and, where appropriate, UHF
reference states. Where available, we also compare with results from
algebraic-diagrammatic construction
(ADC),\cite{schirmer_beyond_1982,trofimov_molecular_2005,QuAcK}
equation-of-motion coupled-cluster
(EOM-CC),\cite{stanton_equation_1993,
christiansen_second-order_1995,epifanovsky_software_2021}
density matrix renormalization group
(DMRG),\cite{chan_density_2011,zhai_block2_2023}
and full configuration interaction (FCI).
CAS($n$,$m$) denotes the zeroth-order results with
$n$ active electrons in $m$ active spatial orbitals, whereas
MR-$GW$($n$,$m$) denotes the corresponding MR-$GW$ results.
Detailed computational protocols and additional numerical results are
provided in the Supplemental Material.\cite{SM}}

\begin{figure}[t]
    \centering
        \includegraphics[width=0.95\linewidth]{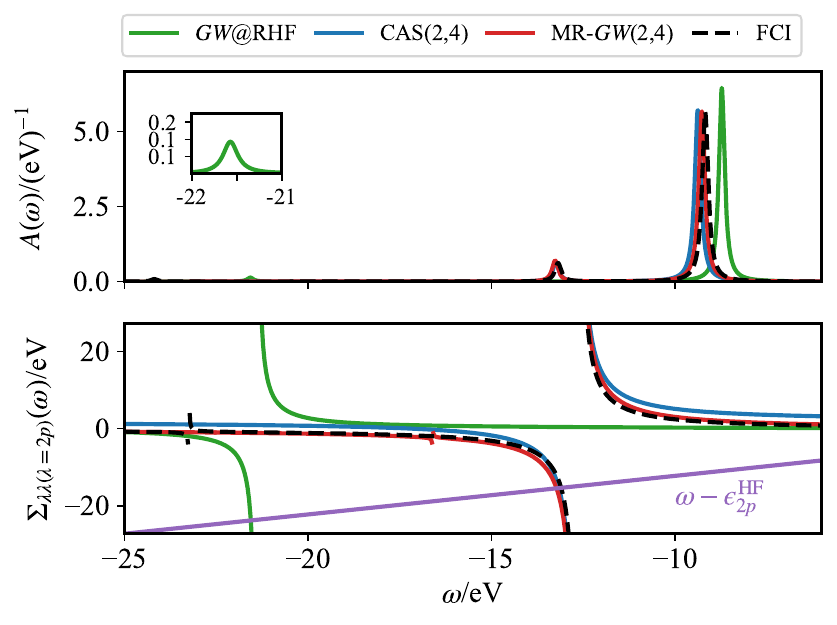}
    \caption{Top panel: spectral functions of \ce{Be} calculated by different methods with a broadening of 0.1 eV. 
    Bottom panel: self-energies on the right-hand side of Eq. \eqref{eq:dyson-diag} and the graphical solution (purple) of Eq. \eqref{eq:dyson-diag}
    for the $2p$ ($\epsilon^{\mathrm{HF}}_{2p}=0.0824$ Hartree) orbital.}
    \label{fig:Be}
\end{figure}

\textit{\ce{Be} atom---}The ground state of the \ce{Be} atom is known to be multi-configurational\cite{krylov_size-consistent_2001,casanova_spin-flip_2020, monino_spin-conserved_2021}, where both the $(1s)^2(2s)^2$ and $(1s)^2(2p)^2$ configurations matter, while
the RHF reference captures only the former. 
The CASSCF reference with a CAS(2,4) active space containing $2s$ and three $2p$ orbitals shows
non-negligible contribution from the $(1s)^2(2p)^2$ configurations, viz.,
$|\Phi_0\rangle = 0.948|(1s)^2(2s)^2\rangle -0.184|(1s)^2(2p_x)^2\rangle-0.184|(1s)^2(2p_y)^2\rangle-0.184|(1s)^2(2p_z)^2\rangle$ with the 6-31G basis set.\cite{hehre_selfconsistent_1972}
Figure \ref{fig:Be} (top panel) shows
the spectral functions calculated by $GW$ and MR-$GW$, 
compared to FCI.
{\color{\mycolor}
While $GW$@RHF reproduces the principal peak near 9 eV, which is dominated by ionization from the $2s$ orbital and denoted $(2s)^{-1}$, it provides a qualitatively incorrect description of the satellite near 13 eV, placing it at a substantially higher energy with a negligible spectral weight. By contrast, both CAS(2,4) and MR-$GW$(2,4) capture the principal and satellite peaks.}

To expose the origin of this satellite, we choose $\mathbf{g}_0$ as the RHF Green's function and use the diagonal approximation to Eq. \eqref{eq:dyson-unified}
\begin{align}
    \omega - \epsilon_{\lambda}^{\mathrm{HF}} = \Re[\Sigma_0(\omega)+
    \Sigma^{\text{MR-}GW}(\omega)]_{\lambda\lambda}, \label{eq:dyson-diag}
\end{align}
where $\lambda$ labels an RHF orbital 
and $\epsilon_{\lambda}^{\mathrm{HF}}$ its orbital energy. 
Plotting both sides of Eq. \eqref{eq:dyson-diag} for 
$\lambda=2p$ (Fig. \ref{fig:Be}, bottom panel) reveals that
this satellite originates from the $2p$ orbitals. 
Compared to the standard $GW$ self-energy, 
the self-energies of CAS(2,4) and MR-$GW$(2,4) 
are in significantly better agreement with the exact FCI self-energy,
due to the inclusion of the $2p$ orbitals in the active space.
This satellite is thus better understood in a multi-configurational picture, where the $(1s)^2(2p)^2$ configuration has a non-negligible weight in the ground state, such that ionization from the $2p$ orbitals occurs at a much lower energy than the standard $GW$ predicts. {\color{\mycolor}
Table S1 of the Supplemental Material\cite{SM} shows that the same trend persists with larger basis sets, and other single-reference Green's-function methods, including ADC(2) and ADC(3), likewise fail to describe the satellite satisfactorily.
}

\begin{figure}[t]
    \centering
        \includegraphics[width=0.95\linewidth]{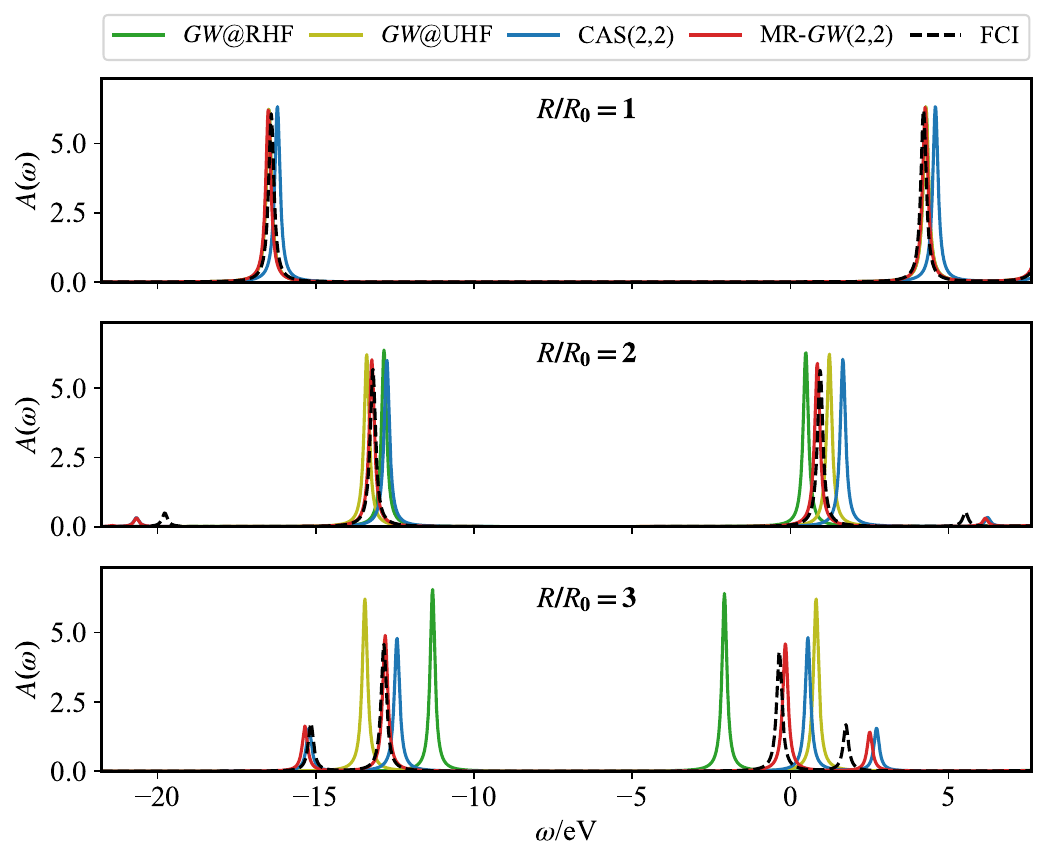}
    \caption{Spectral functions of \ce{H2} calculated by different methods 
    in the cc-pVTZ basis set\cite{dunning_gaussian_1989}
    with a broadening of 0.1 eV, at three representative bond lengths ($R=1R_0$, $2R_0$, and $3R_0$), where $R_0=0.74144~\text{\AA}$ is
    the equilibrium bond distance. {\color{\mycolor}
    The CAS(2,2) and MR-$GW$(2,2) curves use canonical RHF orbitals; CASSCF-orbital results are compared in the Supplemental Material\cite{SM}.}
    }
    \label{fig:h2}
\end{figure}

\textit{Stretched \ce{H2}---}The stretched \ce{H2} is another typical 
strongly correlated molecule\cite{cohen2008insights,caruso_bond_2013}.
Upon bond stretching, the configuration $|\sigma_u^2\rangle$ becomes increasingly important in the ground state
alongside the dominant $|\sigma_g^2\rangle$. 
Fig. \ref{fig:h2} compares the spectral functions 
calculated by different methods at $R=1R_0$, $2R_0$, and $3R_0$, where $R_0=0.74144$ \AA\ is the equilibrium bond length. 
At equilibrium, $GW$@RHF agrees well with FCI for both the first ionization potential (IP) and electron affinity (EA) originating from the $(\sigma_g)^{-1}$ ionization and $(\sigma_u)^{+1}$ attachment, respectively.
{\color{\mycolor}
However, its performance deteriorates quickly upon bond stretching.
Specifically, the errors of the first IP and EA predicted by $GW$@RHF increase, and the satellites emerging near the principal IP and EA peaks,
which originate from the $(\sigma_u)^{-1}$ ionization and $(\sigma_g)^{+1}$ attachment, respectively, are entirely missed (Fig. \ref{fig:h2}).
Broken-symmetry $GW$@UHF improves the principal ionization and attachment peaks over $GW$@RHF upon stretching, but retains sizable errors at $3R_0$ and misses the correlation satellites. By contrast, CAS(2,2) with $\sigma_g$ and $\sigma_u$ as active orbitals already contains the $(\sigma_u)^{-1}$ and $(\sigma_g)^{+1}$ satellite poles, and MR-$GW$ substantially improves the accuracy of CAS(2,2) via
dynamical screening for the leading charged excitations. 
Orbital-dependence tests, including the greater sensitivity for electron attachment, are given in the Supplemental Material\cite{SM}.}

\textit{Ozone---}
The photoelectron spectrum of ozone\cite{cccbdb_nist} (\ce{O3}) has long presented a stringent challenge for
conventional Green's-function methods\cite{decleva1988theoretical,ortiz1998single,mckellar1998complete}
due to its peculiar biradical character\cite{miliordos_unusual_2013}.
In particular, 
the relative ordering among the first three ionized states ($^2A_1$, $^2B_2$, and $^2A_2$) is challenging to predict.  
To describe these three states, a minimal CAS(6,4) active space comprising the doubly occupied $4b_2$, $6a_1$, $1a_2$ RHF orbitals and the lowest virtual orbital ($2b_1$) is required. The CASCI reference state within this active space is dominated by two determinants, e.g.,
$|\Phi_0\rangle = 0.936|(\text{cs})(1a_2)^2\rangle - 0.349 |(\text{cs})(2b_1)^2\rangle$ in the 6-31G basis set, where (cs) denotes the lowest 11 closed-shell orbitals.

\begin{figure}
    \centering
    \includegraphics[width=\linewidth]{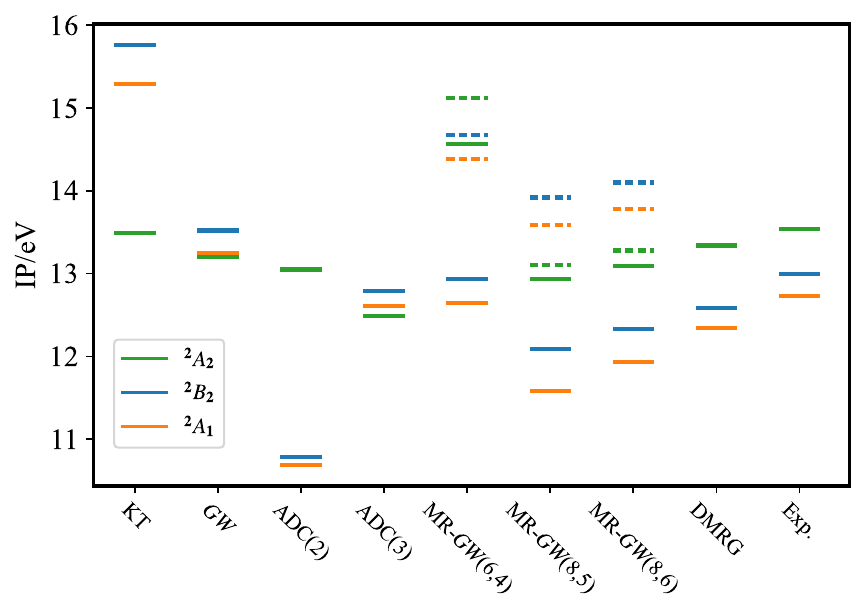} \\
    \caption{Vertical ionization energies (in eV) of \ce{O3} calculated by different methods in the 6-31G basis set.
    The CAS(6,4) minimal active space contains $4b_2, 6a_1, 1a_2$, and $2b_1$ orbitals, while CAS(8,5) additionally contains $1b_1$ and CAS(8,6) contains both $1b_1$ and $7a_1$.
    The corresponding zeroth-order CASCI results with RHF orbitals are plotted in dashed lines.
    Experimental data are taken from Ref. \cite{wiesner_valence_2003}.
    }
    \label{fig:o3}
\end{figure}

Figure~\ref{fig:o3} shows that Koopmans' theorem (KT), $GW$@RHF,
and ADC(3) predict an incorrect ordering of the $^2A_1$, $^2B_2$,
and $^2A_2$ ionized states. In particular, all three methods assign
the first ionization to the $^2A_2$ state, which arises predominantly
from the highest occupied molecular orbital (HOMO), $1a_2$.
ADC(2) recovers the correct ordering only through fortuitous error
cancellation and deviates from the experimental ionization energies
by as much as 2.2~eV.\cite{decleva1988theoretical,
wiesner_valence_2003}
In contrast, MR-$GW$ based on a minimal CAS(6,4) active space restores
the experimental state ordering. Enlarging the active space further
improves the overall quantitative agreement, particularly when the previously
identified $1b_1$ orbital is included.\cite{decleva1988theoretical}
{\color{\mycolor}
Larger-basis tests shift the
absolute ionization energies but preserve the state ordering, see the Supplemental Material\cite{SM} for details.
}

{\color{\mycolor}
The importance of the multireference treatment becomes
more apparent as the
\ce{O-O-O} angle decreases toward $90^\circ$,
where the multiconfigurational character becomes more pronounced.
The bending scan in the Supplemental Material shows that MR-$GW$(8,6) follows the main angular trends of the DMRG ionization energies and substantially improves upon
$GW$@RHF and ADC(2) for the $^2A_1$ and $^2B_2$ states.
EOM-CCSD is accurate near equilibrium, but its $^2A_1$ error reaches
$0.91$~eV at $90^\circ$, compared with $-0.25$~eV for MR-$GW$(8,6).
}

\textit{Conclusion---}{\color{\mycolor}We have established a diagrammatic route from single-reference $GW$ to a multireference Green's-function theory built on an interacting
active-space reference, without relying on a direct generalization of
Hedin's equations to an interacting reference.
The active-space reference treats competing configurations
nonperturbatively, while the residual interaction incorporates
dynamical screening through a selected $GW$-like resummation.}
Standard $GW$ is recovered exactly when the active
space is absent, or when its interaction is switched off and the
corresponding Slater-determinant reference is used.
For the prototypical correlated atoms and molecules considered here,
MR-$GW$ captures many-body satellite structure missed by single-reference $GW$, and generally improves the leading ionization and attachment energies.

The close algebraic correspondence between the resulting MR-$GW$ and
standard $GW$ equations should enable established resolution-of-identity (RI) 
and response techniques\cite{ren_resolution--identity_2012,govoni_large_2015,
zhu_all-electron_2021} to be adapted to MR-$GW$ calculations in large
orbital spaces.
{\color{\mycolor}
For larger active spaces, iterative response algorithms\cite{dorando_analytic_2009,koch_fci_response_1991,
olsen_large_response_1988}
can avoid full diagonalization of the active-space Hamiltonian and
MR-RPA matrix, as outlined in the Supplemental Material.\cite{SM}
More broadly, recent multichannel and particle-particle Green's-function
developments\cite{riva_multichannel_2023,riva_derivation_2024,
marie_anomalous_2024,wang_unified_2025}
further suggest extending MR-$GW$ to other diagrammatic channels.}
Together, these advances open a path toward spectroscopic calculations
for correlated defects\cite{mitra_excited_2021,haldar_local_2023,
benedek_accurate_2025} and $d$- and $f$-electron systems that remain
challenging for single-reference approaches.








\let\oldaddcontentsline\addcontentsline
\renewcommand{\addcontentsline}[3]{}

\section*{Acknowledgment}
The authors acknowledge helpful discussions with Xinguo Ren, Hong Jiang,
and Zhebin Guan. This work was supported by the Quantum Science and Technology-National Science and Technology Major Project (2023ZD0300200) and the Fundamental Research Funds for the Central Universities.

\section*{Data Availability Statement}
{\color{\mycolor}
The MR-$GW$ code developed in this work is publicly available\cite{MRMBPTcode}.
Data reported in this work are provided in Ref. \cite{datarepo}.
}

\bibliographystyle{apsrev4-2}
\bibliography{mrgw}


%
\clearpage
\section*{End Matter}
{\it Appendix A: Details of the Dyall Hamiltonian---}The total Hamiltonian reads
\begin{align}
    \hH = h_{pq}\hp^\dagger\hq + \frac{1}{2}\eriph{pq}{rs}\hp^\dagger\hq^\dagger\hs\hr,
\end{align}
where $h_{pq}$ and $\eriph{pq}{rs}$ are the one-electron and
two-electron integrals, respectively.
For the Dyall Hamiltonian \eqref{eq:Hdyall-main2}, the inactive orbitals $\{P,Q,\cdots\}$ are further partitioned  into doubly-occupied core orbitals $\{i,j,\cdots\}$ and virtual orbitals $\{a,b,\cdots\}$.
The core and virtual orbitals are canonicalized by diagonalizing 
two Fock matrices,
\begin{align}
    F_{ij} =& h_{ij} + \langle ik||jk \rangle + \langle ix||jy\rangle\langle \hat{x}^\dagger\hat{y}\rangle = \epsilon_i \delta_{ij}, \nonumber\\
    F_{ab} =& h_{ab} + \langle ak||bk \rangle + \langle ax||by\rangle\langle \hat{x}^\dagger\hat{y}\rangle = \epsilon_a \delta_{ab}, 
\end{align}
respectively, whose eigenvalues are $\epsilon_P$ \eqref{eq:Hdyall-main2}. Here, $\langle \hat{x}^\dagger \hat{y}\rangle$ denotes the active-space one-body density matrix of the CASCI/CASSCF wavefunction, 
and $\erias{pq}{rs} = \langle pq|rs\rangle - \langle pq|sr\rangle$ are the antisymmetrized two-electron integrals. 
The effective one-electron integrals within the active space $h^\text{eff}_{xy}$ \eqref{eq:Hdyall-main2} are defined as a mean-field generated by only the core electrons
\begin{align}
    h^\text{eff}_{xy} = h_{xy} + \langle xk||yk \rangle.
\end{align}

With $\hH^\text{Dyall}_0$ fully specified, explicit expressions for
the one-electron ($u_{pq}$) and two-electron integrals ($v_{pr,qs}$) for $\hat{V}$ \eqref{eq:PT-V} 
can be readily found via $\hat{V}=\hH - \hH^\text{Dyall}_0$. 
Specifically, $[u_{pq}]$ is 
a Hermitian block matrix
\begin{align}
    \mathbf{u} = \begin{pmatrix}
        [u_{ij}]  & [u_{ib}] & [u_{iy}] \\
        [u_{aj}]  & [u_{ab}] & [u_{ay}] \\
        [u_{xj}]  & [u_{xb}] & [u_{xy}] \\
    \end{pmatrix},
\end{align}
where the lower triangular blocks are 
\begin{align}
    u_{ij} &= h_{ij} - F_{ij} =  -\langle ik||jk \rangle - \langle ix||jy\rangle\langle \hat{x}^\dagger\hat{y}\rangle,\nonumber\\
    u_{aj} &= h_{aj},\nonumber\\
    u_{xj} &= h_{xj},\nonumber\\
    u_{ab} &= h_{ab} - F_{ab} = - \langle ak||bk \rangle - \langle ax||by\rangle \langle\hat{x}^\dagger\hat{y}\rangle, \nonumber\\
    u_{xb} &= h_{xb},\nonumber\\
    u_{xy} &= h_{xy} - h_{xy}^{\mathrm{eff}} = -\erias{xk}{yk}.\label{eq:uxy}
\end{align}
The two-electron perturbation $v_{pr,qs}$ is given by
\begin{align}
    v_{pr,qs} =
    (1-\delta_{p\in A}\delta_{r\in A}\delta_{q\in A}\delta_{s\in A}) \eriph{pq}{rs},\label{eq:V2dyall}
\end{align}
where $\delta_{p\in A}=1(0)$ if $p$ is an (in)active orbital.

{\it Appendix B: First-order self-energy diagrams---}Direct
expansion of the Green's function in the residual interaction,
followed by cumulant decomposition\cite{negele_quantum_1998,metzner1991linked}, 
gives the first-order structure
(see the Supplemental Material\cite{SM})
\begin{align}
    \mathbf{G}_{1} =& \mathbf{G}_0 \mathbf{\Sigma}^{11}_1 \mathbf{G}_0 + \mathbf{G}_0\mathbf{\Sigma}^{12}_1 + \mathbf{\Sigma}^{21}_1 \mathbf{G}_0 + \mathbf{\Sigma}^{22}_1,\label{eq:GDE-1storder}
\end{align}
and identifies $\mathbf{\Sigma}_1^{ij}$ in the time domain, 
\begin{align}
    [\Sigma^{11}_1]_{rs}(t,t^\prime) &= \delta(t-t^\prime)
    u^{\mathrm{eff}}_{rs}
    ,\label{eq:se-GDE-1}\\
    [\Sigma^{12}_1]_{rq}(t,t^\prime)&=-\frac{\ii}{2}\bar{v}_{rt,su}[G_{0}^{c}]_{ut,sq}(t^+,t,t^{++},t^\prime),\label{eq:se-GDE-2}\\
    [\Sigma^{21}_1]_{pu}(t,t^\prime)&=-\frac{\ii}{2}[G_{0}^c]_{pt,rs}(t,t^\prime,t^{\prime ++},t^{\prime +})\bar{v}_{rt,su},\label{eq:se-GDE-3}\\
    [\Sigma^{22}_1]_{pq}(t,t^\prime) &= \int_{-\infty}^{\infty} dt_1\,  
    \left(-u^{\mathrm{eff}}_{rs}[G_0^c]_{ps,qr}(t,t_1,t^\prime,t_1^+)\right.\nonumber\\
    +\frac{\ii}{4}\bar{v}_{rt,su} & \left. [G_0^c]_{ptu,qrs}(t,t_1,t_1^+,t^\prime,t_1^{+++},t_1^{++})
    \right),
    \label{eq:se-GDE-4} 
\end{align}
where 
$G_0^c$ denotes the zeroth-order two- or three-body connected Green's functions, 
$\bar{v}_{pr,qs} = v_{pr,qs} - v_{ps,qr}$ is the antisymmetrized interaction,
and $u^{\mathrm{eff}}_{pq}=
u_{pq}+\bar{v}_{pq,rs}(-\ii)[G_0]_{sr}(t,t^+)
=
u_{pq}+\bar{v}_{pq,rs}\langle \hat{r}^\dagger \hat{s}\rangle$.
Figure \ref{fig:GDE-1} displays the corresponding diagrammatic representations for $\mathbf{\Sigma}^{ij}_1$. 

\begin{figure}[t]
\centering
    \begin{subfigure}{\linewidth}
    \caption{\hspace{\linewidth} }
    \vspace{-8pt}
    \begin{tikzpicture}[scale=.55]
        \draw[very thick] (0,0) circle[radius=.15];
        \node at (2,0) {$+$};
        \gfloop{(4,.5)}{90}{.4}
        \Vline{(4,-1)}{90}{1.5}

        \node at (6,0) {$+$};

        \Vline{(8,0)}{0}{2}

        \gfline[-45]{(8,0)}{0}{2}
    \end{tikzpicture}
    \end{subfigure}

    \begin{subfigure}{.45\linewidth}
    \caption{\hspace{\linewidth} }
    \vspace{-8pt}
    \begin{tikzpicture}[scale=.6, rotate=90]
        \gflinex[0]{(0,0)}{-90}{2}
        \gflinex[40]{(0,2)}{-90}{2}
        \gflinex[0]{(0,2)}{-90}{2}
        \gflinex[40]{(0,0)}{90}{2}
        \Vdot{(0,2)}
        \cmdot[6]{(0,0)}
    \end{tikzpicture}
    \end{subfigure}
    \hfill
    \begin{subfigure}{.45\linewidth}
    \caption{\hspace{\linewidth} }
    \vspace{-8pt}
    \begin{tikzpicture}[scale=.6, rotate=-90]
        \gflinex[0]{(0,-2)}{90}{2}
        \gflinex[40]{(0,2)}{-90}{2}
        \gflinex[0]{(0,0)}{90}{2}
        \gflinex[40]{(0,0)}{90}{2}
        \cmdot[6]{(0,0)}
        \Vdot{(0,2)}
    \end{tikzpicture}
    \end{subfigure}
    
    \begin{subfigure}{\linewidth}
    \caption{\hspace{\linewidth} }
    \vspace{-4pt}
    \begin{tikzpicture}[scale=.6]
    \gflinex[0]{(.5,0)}{0}{1.5}
    \gflinex[0]{(2,0)}{0}{1.5}
    \phbblx{(2,0)}{90}{2}
    \draw[very thick] (2,2) circle[radius=.15];
    \cmdot[6]{(2,0)}

    \node at (4.5,0) {$+$};

    \gflinex[0]{(0+5.5,0)}{0}{1.5}
    \gflinex[0]{(2+5,0)}{0}{1.5}
    \phbblx{(2+5,0)}{90}{1.2}
    \gfloop{(2+5,1.2)}{90}{.4}
    \Vdot{(2+5,1.2)}
    \cmdot[6]{(2+5,0)}

    \node at (9.5,0) {$+$};

    \gflinex[0]{(0+10.5,0)}{0}{1.5}
    \gflinex[0]{(1.5+10.5,0)}{0}{1.5}
    \phbblx[20]{(12,0)}{90}{2}
    \phbblx[70]{(12,0)}{90}{2}
    \cmdot[6]{(12,0)}
    \Vdot{(12,2)}
    
    \end{tikzpicture}
    \end{subfigure}

\caption{Diagrammatic representation of first-order self-energies $\mathbf{\Sigma}_1^{ij}$. 
(a) $\mathbf{\Sigma}^{11}_1$: one-body $u_{pq}$, Hartree, and exchange self-energies; (b) $\mathbf{\Sigma}^{12}_1$; (c) $\mathbf{\Sigma}^{21}_1$; (d) $\mathbf{\Sigma}^{22}_1$.
The black directed lines represent $G_0$, and
the wiggly lines represent two-electron interactions $v_{pr,qs}$. The open and filled circles represent the one-electron $u_{pq}$ and antisymmetrized two-electron interactions $\bar{v}_{pr,qs}$, respectively.
The red squares with directed lines represent zeroth-order connected two-body or three-body Green's functions $G_0^c$.}
\label{fig:GDE-1}
\end{figure}

{\color{\mycolor}
\textit{Appendix C: Impact of $\mathbf{\Sigma}_{1}^{12}$ and
$\mathbf{\Sigma}_{1}^{21}$ on the spectral function---}
The use of the Dyall Hamiltonian leads to the additional simplification
$\mathbf{\Sigma}_{1}^{22}=0$ (see the Supplemental
Material\cite{SM} for details).
Figure~\ref{fig:allblock-diagnostic} illustrates the effects of the
mixed self-energy blocks $\mathbf{\Sigma}_{1}^{12/21}$ 
on the spectral functions of \ce{Be} and
linear \ce{H4}. For the systems examined, including these blocks
produces only minor changes in the peak positions obtained with either
CAS or MR-$GW$. Their isolated inclusion can,
however, violate the positive definiteness of $A(\omega)$, as
illustrated by \ce{H4}. As detailed in the Supplemental
Material,\cite{SM} this behavior originates from the higher-order poles introduced into $\mathbf{M}(\omega)$ by this incomplete treatment. Given their negligible effect on the peak positions at the present order and the unphysical analytical structure generated when they are included without the complementary higher-order contributions, we omit $\mathbf{\Sigma}_{1}^{12/21}$ from
the present definition of MR-$GW$. This choice also preserves a direct
diagrammatic correspondence between MR-$GW$ and standard $GW$.
A consistent treatment of these mixed blocks will require an
order-consistent higher-order completion or resummation; recent ADC
reformulations of vertex-corrected $GW$ that enforce a
positive-semidefinite self-energy structure may provide useful
guidance.\cite{marie_adc_gw_2026}}

\begin{figure}[t]
    \centering
\includegraphics[width=.95\linewidth]{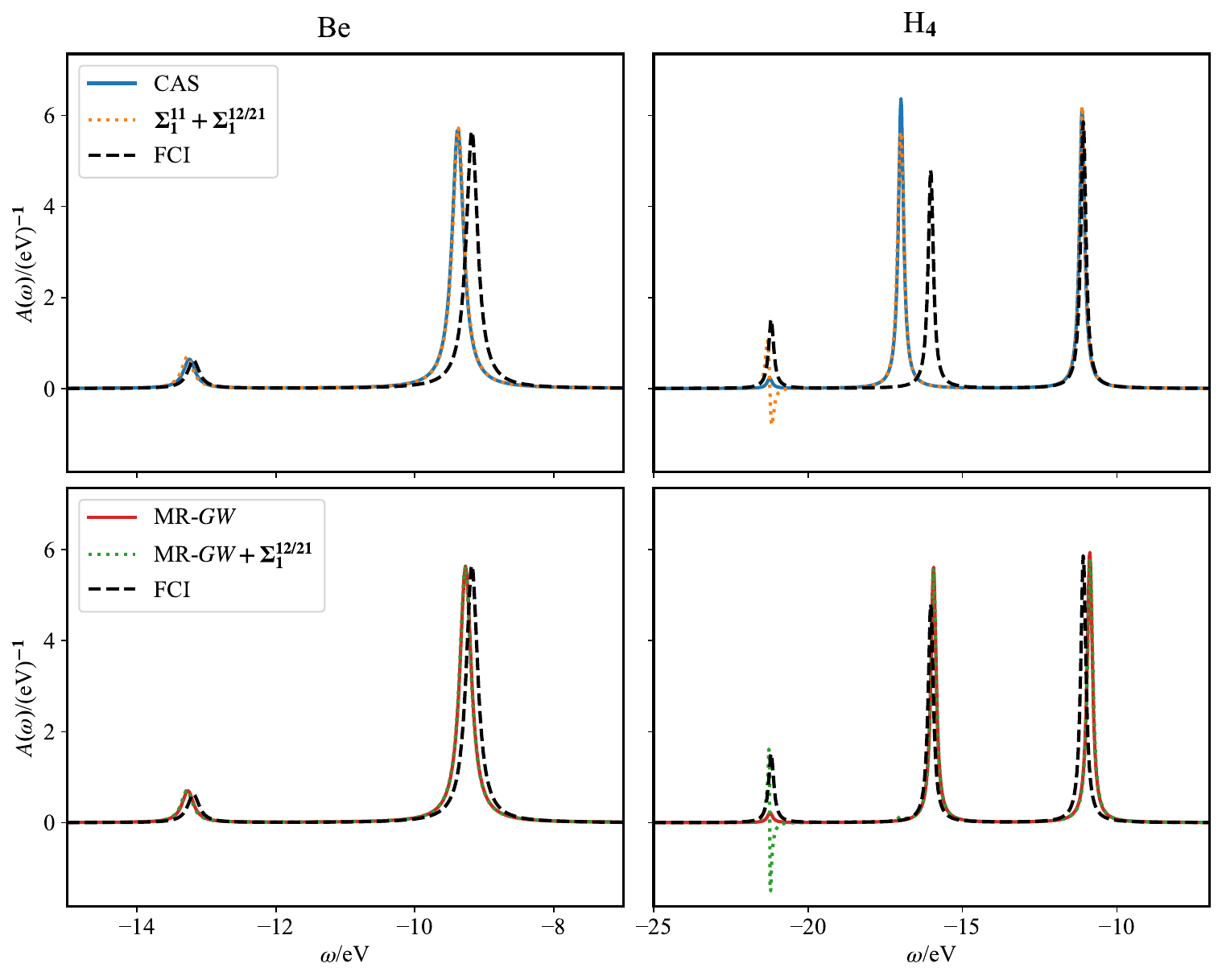}
\captionof{figure}{{\color{\mycolor}
Spectral functions of Be and linear \ce{H4} 
obtained with and without the mixed self-energy blocks $
\mathbf{\Sigma}_{1}^{12/21}$ using a broadening of 0.1 eV.
Calculations for \ce{Be} use the same setting
as in Fig. \ref{fig:Be}, while
calculations for \ce{H4} with \ce{H} atoms separated by 1 \AA~
employ the STO-6G basis set and a CAS(2,2) active space containing the 
CASSCF HOMO and LUMO orbitals. FCI spectral functions (black)
are also shown for comparison.}}
\label{fig:allblock-diagnostic}
\end{figure}

{\color{\mycolor}
\textit{Appendix D: Computational workflow and cost---}
The present pilot implementation uses an explicit spectral
representation. FCI diagonalization of the Dyall Hamiltonian
[Eq.~\eqref{eq:Hdyall-main2}] supplies $\mathbf{G}_0$ and the
active-space transition densities; dense MR-RPA diagonalization
[Eq.~\eqref{eq:MRRPAdiag}] then yields the screened-interaction poles
used in the self-energy [Eq.~\eqref{eq:sigma-mrgw-general}], followed
by solution of the Dyson equation [Eq.~\eqref{eq:dyson-mrgw}].
Let $N_{\rm orb}=N_c+N_a+N_v$, where
$N_{c}$, $N_a$, and $N_v$ denote the number of core, active,
and virtual orbitals, respectively, $D$ be the largest active-sector determinant
dimension, and $K$ be the dimension of each
MR-RPA $\mathbf A$ or $\mathbf B$ block, with
$K\leq N_cN_v+D(N_c+N_v+1)$ (the full matrix is $2K\times2K$).\cite{wang_generalized_2025}
The principal costs of the present implementation 
are $\mathcal{O}(D^3)$ and $\mathcal{O}(K^3)$ for diagonalizing
the active Hamiltonian and MR-RPA, respectively.
Without an active space, $K=\mathcal{O}(N_{\rm orb}^2)$ and the method
reduces to a dense spectral $G_0W_0$ implementation with
$\mathcal{O}(N_{\rm orb}^6)$ scaling.\cite{van_setten_gw100_2015} At fixed
active space with $N_a \ll N_c$ and $N_v$, 
the present implementation has the same orbital-space
scaling as the corresponding $G_0W_0$ algorithm, plus
determinant-space costs.

Iterative frequency-domain response can remove the dense
diagonalization and explicit sum over states,\cite{dorando_analytic_2009,koch_fci_response_1991,
olsen_large_response_1988}. To also reduce the orbital-space scaling, it must be combined with RI or related modern $G_0W_0$ techniques, which require $\mathcal{O}(N_{\rm orb}^4)$ work for selected quasiparticle energies and can reach $\mathcal{O}(N_{\rm orb}^3)$ in space--time formulations.\cite{ren_resolution--identity_2012,
wilhelm_low-scaling_2021,peng_basis-set-error-free_2023} The step-by-step costs of the present implementation and this prospective route are detailed in the Supplemental Material.\cite{SM}
}

\let\addcontentsline\oldaddcontentsline

\raggedbottom
\clearpage
\widetext
\allowdisplaybreaks[4]


\begin{center}
{\color{\mycolor}\textbf{\large Supplemental Material for \\
``A Diagrammatic Route to Multi-Reference $GW$ for Strongly Correlated Molecules"}}\\
\vspace{2ex}
Yuqi Wang$^{1,2}$, Wei-Hai Fang$^{1,2}$, and Zhendong Li$^{1,2,*}$ \\
{\it $^1$ Key Laboratory of Theoretical and Computational Photochemistry, Ministry of Education, College of Chemistry, Beijing Normal University, Beijing, 100875, China \\
$^2$ Institute for Advanced Study, Beijing Normal University, Beijing, 100875, China}
\end{center}

\setcounter{secnumdepth}{3}
\setcounter{section}{0}
\setcounter{equation}{0}
\setcounter{figure}{0}
\setcounter{table}{0}
\setcounter{page}{1}
\makeatletter
\renewcommand{\thesection}{S\arabic{section}}
\renewcommand{\thefigure}{S\arabic{figure}}
\renewcommand{\thetable}{S\arabic{table}}
\renewcommand{\appendixname}{}
\counterwithout{equation}{section} 
\renewcommand{\theequation}{S\arabic{equation}}
\tableofcontents

\section{First-order perturbation for the Generalized Dyson equation}
To obtain the first-order self-energies ($\mathbf{\Sigma}_1^{ij}$) in the generalized Dyson equation \eqref{eq:GDE-11}, 
we first derive the general expressions for a perturbation in the form of Eq. \eqref{eq:PT-V}
without assuming $\hH_0$ is the Dyall Hamiltonian. 
Further simplifications with the Dyall Hamiltonian are given in Sec. \ref{eq:CASCI-1st-SE}.

\subsection{Cumulant decomposition of time-ordered Green's functions}
We recapitulate 
the cumulant decomposition\cite{negele_quantum_1998,metzner1991linked} for one-body, two-body, and three-body Green's functions, which will be used in the subsequent sections. 
For simplicity, $\hat{p}^{(\dagger)}(t_1)$ in the Heisenberg representation is abbreviated as $\hp^{(\dagger)}_1$.

\textit{One-body Green's function:} 
\begin{align}
    G_{pq}(t_1,t_2) \equiv
    (-\ii)\langle \mathcal{T}[\hp_1\hq^\dagger_2]\rangle = G^c_{pq}(t_1,t_2) ,\label{eq:cml-1e}
\end{align}

\textit{Two-body Green's function:}
\begin{align}
G_{sr,pq}(t_3,t_4,t_2,t_1) &\equiv (-\ii)^2\langle \mathcal{T}[\hs_3\hr_4 \hq^\dagger_1\hp^\dagger_2]\rangle \nonumber\\
&= G^c_{sr,pq}(t_3,t_4,t_2,t_1)- G_{sq}(t_3,t_1)G_{rp}(t_4,t_2)+ G_{rq}(t_4,t_1)G_{sp}(t_3,t_2), \label{eq:cml-2e}
\end{align}

\textit{Three-body Green's function:}

\begin{align}
G_{stu,rqp}(t_4,t_5,t_6,t_3,t_2,t_1) & \equiv
    (-\ii)^3\langle \mathcal{T}[\hs_4\hat{t}_5\hat{u}_6\hp^\dagger_1\hq^\dagger_2\hr^\dagger_3]\rangle \nonumber\\
    &= G^c_{stu,rqp}(t_4,t_5,t_6,t_3,t_2,t_1)\nonumber\\
    &+ G_{sp}(t_4,t_1) G^c_{tu,rq}(t_5,t_6,t_3,t_2) 
    - G_{tp}(t_5,t_1) G^c_{su,rq}(t_4,t_6,t_3,t_2)
    + G_{up}(t_6,t_1) G^c_{st,rq}(t_4,t_5,t_3,t_2)\nonumber\\
    &- G_{sq}(t_4,t_2) G^c_{tu,rp}(t_5,t_6,t_3,t_1)
    + G_{tq}(t_5,t_2)G^c_{su,rp}(t_4,t_6,t_3,t_1)
    - G_{uq}(t_6,t_2)G^c_{st,rp}(t_4,t_5,t_3,t_1)\nonumber\\
    &+ G_{sr}(t_4,t_3) G^c_{tu,qp}(t_5,t_6,t_2,t_1) 
    - G_{tr}(t_5,t_3) G^c_{su,qp}(t_4,t_6,t_2,t_1) 
    + G_{ur}(t_6,t_3) G^c_{st,qp}(t_4,t_5,t_2,t_1) \nonumber\\
    & - G_{sp}(t_4,t_1)G_{tq}(t_5,t_2)G_{ur}(t_6,t_3) 
    - G_{tp}(t_5,t_1)G_{uq}(t_6,t_2)G_{sr}(t_4,t_3)\nonumber\\
    &- G_{up}(t_6,t_1)G_{sq}(t_4,t_2)G_{tr}(t_5,t_3) 
    + G_{sp}(t_4,t_1)G_{uq}(t_6,t_2)G_{tr}(t_5,t_3)\nonumber\\
    &+ G_{tp}(t_5,t_1)G_{sq}(t_4,t_2)G_{ur}(t_6,t_3) 
    + G_{up}(t_6,t_1)G_{tq}(t_5,t_2)G_{sr}(t_4,t_3),\label{eq:cml-3e}
\end{align}

These results apply to both interacting and noninteracting Green's functions.


\subsection{First-order Green's function and self-energies}
The perturbation expansion for the interacting Green's function by the Gell-Mann–Low theorem\cite{negele_quantum_1998} reads,
\begin{align}
G_{pq}(t,t^{\prime}) &=\lim_{\eta\rightarrow 0^+}\frac{(-\ii)\langle \Phi_0|\mathcal{T}[\exp(-\ii\int_{-\infty}^{\infty}  dt_1 e^{-\eta|t_1|}\hat{V}(t_1)) \hp(t)\hq^\dagger(t^\prime)]|\Phi_0\rangle}
{\langle \Phi_0|\mathcal{T}[\exp(-\ii\int_{-\infty}^{\infty} dt_1 e^{-\eta|t_1|}\hat{V}(t_1))]|\Phi_0\rangle}\nonumber\\
&=\lim_{\eta\rightarrow 0^+}\frac{(-\ii)\left(\langle\Phi_0|\Tau{\hp(t)\hq^{\dagger}(t^{\prime})}|\Phi_0\rangle+(-\ii)\int_{-\infty}^{\infty} dt_1 e^{-\eta|t_1|}\langle \Phi_0|\Tau{\hat{V}(t_1)\hp(t)\hq^{\dagger}(t^{\prime})}|\Phi_0\rangle+\cdots\right)}{1 + (-\ii)\int_{-\infty}^{\infty} dt_1e^{-\eta|t_1|}\langle\Phi_0|\hat{V}(t_1)|\Phi_0\rangle + \cdots}\nonumber\\
&
\equiv [G_0]_{pq}+[G_1]_{pq}+[G_2]_{pq}+\cdots,\label{eq:GML-G}
\end{align}
where operators are in the interaction picture generated by the interacting $\hat{H}_0$.
We assume that the chosen reference state is adiabatically connected to the target state within the specified symmetry sector; a degenerate
reference requires an explicit state-following or model-space prescription.
Eq. \eqref{eq:GML-G} is of a form
\begin{align*}
\frac{A_0+A_1+A_2+\cdots}{1+B_1+B_2+\cdots}= G_0 +G_1+G_2+\cdots ,
\end{align*}
which can be reorganized to express $G$ at each order as
\begin{align*}
G_{0} =& A_0,\\
G_1 =& A_1-G_0B_1,\\
G_2 =& A_2-G_0B_2-G_1B_1,
\end{align*}
etc. Specifically, for the perturbation $\hat{V}$ given in Eq. \eqref{eq:PT-V},
we can find the first-order Green's function $G_{1,1e}$ due to the one-electron perturbation as
\begin{align}
    [G_{1,1e}]_{pq}(t,t^\prime) =u_{rs}\int_{-\infty}^{\infty}dt_1 
    \left([G_0]_{pr}(t,t_1)[G_0]_{sq}(t_1,t^\prime) - [G_0^c]_{sp,rq}(t_1,t,t_1^+,t^\prime)\right),
\end{align}
and that $G_{1,2e}$ due to two-electron perturbation as
\begin{align}
    [G_{1,2e}]_{pq}(t,t^\prime) &=\frac{\ii}{4}\bar{v}_{rt,su}\int_{-\infty}^{\infty}dt_1\, 
    \left([G_0^c]_{utp,qsr}(t_1^+,t_1,t,t^\prime,t_1^{++},t_1^{+++})\right.\nonumber\\
    &-2[G_0]_{pr}(t,t_1)[G_0^c]_{ut,sq}(t_1^+,t_1,t_1^{++},t^\prime)\nonumber\\
    &-2[G_0]_{uq}(t_1,t^\prime)[G_0^c]_{tp,sr}(t_1,t,t_1^+,t_1^{++})\nonumber\\
    &-4[G_0]_{tr}(t_1,t_1^+)[G_0^c]_{up,qs}(t_1,t,t^\prime,t_1^+)\nonumber\\
    &\left.-4[G_0]_{pr}(t,t_1)[G_0]_{us}(t_1,t_1^+)[G_0]_{tq}(t_1,t^\prime)\right),
\end{align}
where the cumulant decomposition \eqref{eq:cml-1e}-\eqref{eq:cml-3e} has been applied. Combining $G_{1,1e}$ and $G_{1,2e}$, we obtain the first-order Green's function as
\begin{align}
    [G_1]_{pq}(t,t^\prime) &= \int_{-\infty}^{\infty} dt_1\,
\Big\{    
    [G_0]_{pr}(t,t_1)\left(u_{rs}+\bar{v}_{rs,tu}(-\ii)
    [G_0]_{ut}(t_1,t_1^+)
    \right)[G_0]_{sq}(t_1,t^\prime) \nonumber\\
     &+[G_0]_{pr}(t,t_1)\left(-\frac{\ii}{2}\bar{v}_{rt,su}[G_0^c]_{ut,sq}(t_1^+,t_1,t_1^{++},t^\prime)\right)\nonumber\\
     &+\left(-\frac{\ii}{2}[G_0^c]_{pt,rs}(t,t_1,t_1^{++},t_1^+)\bar{v}_{rt,su}\right)[G_0]_{uq}(t_1,t^\prime)\nonumber\\
     &-\left(u_{rs}+\bar{v}_{rs,tu}(-\ii)[G_0]_{ut}(t_1,t_1^+)\right)[G_0^c]_{ps,qr}(t,t_1,t^\prime,t_1^+)\nonumber\\
     &+\frac{\ii}{4}\bar{v}_{rt,su}[G_0^c]_{ptu,qrs}(t,t_1,t_1^+,t^\prime,t_1^{+++},t_1^{++})
\Big\}.
\end{align}
Comparing this result with the first-order 
generalized Dyson equation \eqref{eq:GDE-1storder} allows us to identify
$\Sigma^{ij}_1$ \eqref{eq:se-GDE-1}-\eqref{eq:se-GDE-4}.

\subsection{Simplifications for $\hat{H}_0^{\mathrm{Dyall}}$ and numerical implementation}\label{eq:CASCI-1st-SE}
With $\hH^\text{Dyall}$ as $\hH_0$, further simplifications for $\Sigma^{ij}_1$ can be made:

(1) $\mathbf{\Sigma}_1^{22}(\omega) = 0$, because the first and second diagrams of Fig. \ref{fig:GDE-1} cancel each other out
following Eq. \eqref{eq:uxy}, while the third diagram vanishes because the two-electron interactions involving all active indices are zero, see Eq. \eqref{eq:V2dyall}.

(2) $[\Sigma^{11}_1]_{rs}(\omega) = u_{rs}+\bar{v}_{rs,pq}(-\ii)[G_0]_{qp}(t,t^+) = u_{rs} + \bar{v}_{rs,pq}\langle \hat{p}^\dagger  \hat{q}\rangle$ is frequency-independent and Hermitian, with the following independent matrix elements
\begin{align}
\protect[\Sigma^{11}_1]_{ij} &= 0, \\
\protect[\Sigma^{11}_1]_{aj} &= h_{aj} + \bar{v}_{aj,pq}\langle \hat{p}^\dagger  \hat{q} \rangle, \\
\protect[\Sigma^{11}_1]_{xj} &= h_{xj} + \bar{v}_{xj,pq}\langle \hat{p}^\dagger  \hat{q} \rangle, \\
\protect[\Sigma^{11}_1]_{ab} &= 0, \\
\protect[\Sigma^{11}_1]_{xb} &= h_{xb} + \bar{v}_{xb,pq}\langle \hat{p}^\dagger  \hat{q} \rangle, \\
\protect[\Sigma^{11}_1]_{xy} &= 0.
\end{align}

(3) $\mathbf{\Sigma}^{12}_1(\omega)$ and $\mathbf{\Sigma}^{21}_1(\omega)$ are frequency-dependent. When orbitals are partitioned into the inactive ($I$) and active ($A$) subsets, they have the following block structure
\begin{align}
    \boldsymbol{\Sigma}^{21}_1(\omega)=\begin{bmatrix}
        \mathbf{0} & \quad  \mathbf{0} \\
        \boldsymbol{\Sigma}_1^{21,AI}(\omega) & \quad \mathbf{0} 
    \end{bmatrix},\quad 
    \boldsymbol{\Sigma}^{12}_1(\omega)=\begin{bmatrix}
        \mathbf{0} & \quad \boldsymbol{\Sigma}^{12,IA}_1(\omega)\\
        \mathbf{0} & \mathbf{0}
    \end{bmatrix},
\end{align}
with matrix elements given by
\begin{align}
    \left[\Sigma^{12,IA}_1\right]_{Px}(\omega) &= \frac{1}{2}\bar{v}_{Pz,yw}\left[\frac{\langle \Xi_{0,N_a}^{A}|\hat{y}^\dagger \hat{w} \hat{z} |\Xi_{\mu,N_a+1}^{A}\rangle\langle \Xi_{\mu,N_a+1}^{A}|\hat{x}^\dagger|\Xi_{0,N_a}^{A}\rangle}{\omega-\omega_{\mu,N_a+1}^{A}+\ii 0^+} \right.\nonumber\\
    &\quad+ \left.\frac{\langle \Xi_{0,N_a}^{A}|\hat{x}^\dagger|\Xi_{\mu,N_a-1}^{A}\rangle\langle\Xi_{\mu,N_a-1}^{A}| \hat{y}^\dagger \hat{w}\hat{z} |\Xi_{0,N_a}^{A}\rangle}{\omega+\omega_{\mu,N_a-1}^{A}-\ii 0^+}\right] -\bar{v}_{Pz,yw}\langle \hat{y}^\dagger \hat{w}\rangle  [G^A_0]_{zx}(\omega),
    \label{eq:sigma12IA}
\end{align}
and
\begin{align}
    \left[\Sigma^{21,AI}_1\right]_{xP}(\omega) &= -\frac{1}{2}\bar{v}_{zP,wy}\left[\frac{\langle \Xi_{0,N_a}^{A}|\hat{x}|\Xi_{\mu,N_a+1}^{A}\rangle
    \langle \Xi_{\mu,N_a+1}^{A}|\hat{z}^\dagger \hat{w}^\dagger \hat{y}|\Xi_{0,N_a}^{A}\rangle}{\omega-\omega_{\mu,N_a+1}^{A}+\ii 0^+} \right.\nonumber\\
    &\quad \left.+ \frac{\langle \Xi_{\mu,N_a-1}^{A}|\hat{x}|\Xi_{0,N_a}^{A}\rangle\langle \Xi_{0,N_a}^{A} |\hat{z}^\dagger \hat{w}^\dagger \hat{y} |\Xi_{\mu,N_a-1}^{A} \rangle}{\omega+\omega_{\mu,N_a-1}^{A}-\ii 0^+}\right] +\bar{v}_{zP,wy}\langle \hat{w}^\dagger \hat{y}\rangle  [G^A_0]_{xz}(\omega),\label{eq:sigma21AI}
\end{align}
where $|\Xi^A_{\mu,N_a}\rangle$ is an eigenvector of $\hH^{\text{act}}_0$
with $N_a$ electrons
\begin{align}
    \hH^{\text{act}}_0 |\Xi^A_{\mu,N_a}\rangle = E^A_{\mu,N_a}|\Xi^A_{\mu,N_a}\rangle,\label{eq:active-eigen}
\end{align}
and $\omega^A_{\mu,N_a\pm 1} = E^A_{\mu,N_a\pm 1} - E^A_{0,N_a}$ 
is the zeroth-order charged excitation energy.


The multi-reference generalization of the first-order Green's function approach (MR-GF1), which retains only first-order self-energy blocks, can be defined by
\begin{align}
    \mathbf{G}(\omega) &= \left(\mathbf{M}(\omega)^{-1} - \boldsymbol{\Sigma}^{11}_1(\omega)\right)^{-1},\label{eq:GDE-G1}\\
    \mathbf{M}(\omega) &= [\mathbf{I}-\boldsymbol{\Sigma}^{21}_1(\omega)]^{-1}\left[\mathbf{G}_0(\omega)+\boldsymbol{\Sigma}^{22}_1(\omega)\right][\mathbf{I}-\boldsymbol{\Sigma}^{12}_1(\omega)]^{-1}.\label{eq:GDE-M}
\end{align}
With CASCI/CASSCF reference, Eq. \eqref{eq:GDE-M} is written more explicitly as
\begin{align}
    \mathbf{M}&= \begin{bmatrix}
        \mathbf{I} & \mathbf{0} \\
        -\boldsymbol{\Sigma}^{21,AI}_1 & \mathbf{I} 
    \end{bmatrix}^{-1}\begin{bmatrix}
        \mathbf{G}_{0}^{I} & \mathbf{0} \\
        \mathbf{0} & \mathbf{G}_{0}^{A}
    \end{bmatrix}\begin{bmatrix}
        \mathbf{I} & -\boldsymbol{\Sigma}^{12,IA}_1\\
        \mathbf{0} & \mathbf{I}
    \end{bmatrix}^{-1}\nonumber\\
    &= \begin{bmatrix}
        \mathbf{I} & \mathbf{0} \\
        \boldsymbol{\Sigma}^{21,AI}_1 & \mathbf{I} 
    \end{bmatrix}\begin{bmatrix}
        \mathbf{G}_{0}^{I} & \mathbf{0} \\
        \mathbf{0} & \mathbf{G}_{0}^{A}
    \end{bmatrix}\begin{bmatrix}
        \mathbf{I} & \boldsymbol{\Sigma}^{12,IA}_1\\
        \mathbf{0} & \mathbf{I}
    \end{bmatrix}\nonumber\\
    &= \begin{bmatrix}
        \mathbf{G}_0^I & \mathbf{G}_0^I\boldsymbol{\Sigma}^{12,IA}_1\\
        \boldsymbol{\Sigma}^{21,AI}_1\mathbf{G}_0^I & \quad\boldsymbol{\Sigma}^{21,AI}_1\mathbf{G}_0^I\boldsymbol{\Sigma}^{12,IA}_1 + \mathbf{G}_0^A
    \end{bmatrix},
    \label{eq:M1cas}
\end{align}
where the frequency argument has been omitted for clarity.
{\color{\mycolor}Since $\boldsymbol{\Sigma}^{21,AI}_1$ and $\boldsymbol{\Sigma}^{12,IA}_1$ have the same poles as $\mathbf{G}^A_0$, see 
Eqs. \eqref{eq:sigma12IA} and \eqref{eq:sigma21AI}, 
Eq. \eqref{eq:M1cas} reveals that the term $\boldsymbol{\Sigma}^{21,AI}_1\mathbf{G}_0^I\boldsymbol{\Sigma}^{12,IA}_1$ may cause incorrect analytical structure in the final Green's function,
as demonstrated in Fig. \ref{fig:allblock-diagnostic}.}

\section{Derivations of the MR-$GW$ self-energy}
We give the detailed derivations of the MR-$GW$ self-energy.
Since it shares analogous $GW$ diagrams as the standard $GW$,
our derivation unifies
the standard $GW$ and MR-$GW$, 
which includes three steps: derivations of polarizability, screened interaction, and self-energy.
As in the above section, we first derive the general formulation for a general $\hH_0$, 
and consider the simplifications for $\hH_0^{\mathrm{Dyall}}$ later.

\subsection{Polarizability $\mathbf{\Pi}(\omega)$}
In Ref. \cite{wang_generalized_2025}, we derived the polarizability at the MR-RPA level. 
Here, we 
briefly recapitulate the final results necessary for defining screened
interaction. 
To evaluate $\mathbf{\Pi}(\omega)$ \eqref{eq:Pi} with $\mathbf{\Pi}_0(\omega)$ \eqref{eq:mr-rpa-pi0},
we express it explicitly as
\begin{align}
    \Pi_{pr,qs}(\omega) &\equiv [\Pi_0]_{pr,qs}(\omega) + [\mathbf{\Pi}_0(\omega)\mathbf{v}\mathbf{\Pi}(\omega)]_{pr,qs}\nonumber\\
    &=[\Pi_0]_{pr,qs}(\omega) + \left[\mathbf{\Pi}_0(\omega)\mathbf{v}\mathbf{\Pi}_0(\omega)\right]_{pr,qs} + \left[\mathbf{\Pi}_0(\omega)\mathbf{v}\mathbf{\Pi}_0(\omega)\mathbf{v}\mathbf{\Pi}_0(\omega)\right]_{pr,qs} + \cdots. \label{eq:pi-rpa}
\end{align}
The key observation to evaluate this sum\cite{wang_generalized_2025} 
is to introduce the following auxiliary matrices 
\begin{align}
    v^L_{pr,\mu} &\equiv v_{pr,qs}\begin{bmatrix}
        \langle \Phi_0^N|\hq^\dagger \hs|\Phi_\mu^N\rangle &
\langle \Phi_\mu^N|\hq^\dagger \hs|\Phi_0^N\rangle 
    \end{bmatrix},\\
    v^R_{\mu,qs} &\equiv \begin{bmatrix}
        \langle \Phi_\mu^N|\hp^\dagger \hr|\Phi_0^N\rangle \\
        \langle \Phi_0^N|\hp^\dagger \hr|\Phi_\mu^N\rangle 
    \end{bmatrix}v_{pr,qs},\\
    V_{\mu\nu} &=\begin{bmatrix}
    \langle \Phi_\mu^N|p^\dagger r|\Phi_0^N\rangle 
    v_{pr,qs}
    \langle \Phi_0^N|q^\dagger s|\Phi_\nu^N\rangle &
    \langle \Phi_\mu^N|p^\dagger r|\Phi_0^N\rangle 
    v_{pr,qs}
    \langle \Phi_\nu^N|q^\dagger s|\Phi_0^N\rangle  \\
    \langle \Phi_0^N|p^\dagger r|\Phi_\mu^N\rangle
    v_{pr,qs}
    \langle \Phi_0^N|q^\dagger s|\Phi_\nu^N\rangle &
    \langle \Phi_0^N|p^\dagger r|\Phi_\mu^N\rangle
    v_{pr,qs}
    \langle \Phi_\nu^N|q^\dagger s|\Phi_0^N\rangle  \\
\end{bmatrix},\\
    [D_0]_{\mu\nu}(\omega) &= \begin{bmatrix}
    \delta_{\mu\nu}(\omega-\omega_\mu^N+\ii0^+)^{-1} & 0\\
    0 &-\delta_{\mu\nu}(\omega+\omega_\mu^N-\ii0^+)^{-1}
\end{bmatrix},
\end{align}
such that Eq. \eqref{eq:pi-rpa} can be rewritten as
\begin{align}
    \Pi_{pr,qs}(\omega) =& [\Pi_0]_{pr,qs}(\omega) + \left[\mathbf{\Pi}_0(\omega)\mathbf{v}\mathbf{\Pi}_0(\omega)\right]_{pr,qs}
    + \left[\mathbf{\Pi}_0(\omega)\mathbf{v}^L\mathbf{D}_0(\omega)\mathbf{v}^R\mathbf{\Pi}_0(\omega)\right]_{pr,qs} \nonumber\\
    &+\left[\mathbf{\Pi}_0(\omega)\mathbf{v}^L\mathbf{D}_0(\omega)\mathbf{V}\mathbf{D}_0(\omega)\mathbf{v}^R\mathbf{\Pi}_0(\omega)\right]_{pr,qs}+ \cdots,\nonumber\\
    =&[\Pi_0]_{pr,qs}(\omega) + \left[\mathbf{\Pi}_0(\omega)\mathbf{v}\mathbf{\Pi}_0(\omega)\right]_{pr,qs}+\left[\mathbf{\Pi}_0(\omega)\mathbf{v}^L\mathbf{D}(\omega)\mathbf{v}^R\mathbf{\Pi}_0(\omega)\right]_{pr,qs},
\end{align}
where in the last line we have introduced
\begin{align}
    \mathbf{D}(\omega) \equiv \mathbf{D}_0(\omega)+\mathbf{D}_0(\omega)\mathbf{V}\mathbf{D}(\omega).\label{eq:D-rpa}
\end{align}
Assuming a stable MR-RPA solution with real positive frequencies and eigenvectors normalized in the RPA metric, Eq. \eqref{eq:D-rpa} can be solved analytically\cite{wang_generalized_2025} as
\begin{align}
    \mathbf{D}(\omega) = -\sum_{I>0} \left(
\frac{\mathbf{z}_I\mathbf{z}_I^{\dagger}}{\Omega_I - \omega-\ii0^+}+\frac{\mathbf{z}_{-I}\mathbf{z}_{-I}^{\dagger}}{\Omega_I + \omega-\ii0^+}\right),\label{eq:MRRPAdiag}
\end{align}
by diagonalizing an MR-RPA equation
\begin{align}
\begin{bmatrix}
    \mathbf{A} & \mathbf{B}\\
    \mathbf{B}^* & \mathbf{A}^*
\end{bmatrix}\begin{bmatrix}
    \mathbf{X}_I \\ \mathbf{Y}_I
\end{bmatrix} =& \begin{bmatrix}
    \mathbf{I} & \mathbf{0}\\
    \mathbf{0} & -\mathbf{I}
\end{bmatrix}\begin{bmatrix}
    \mathbf{X}_I \\ \mathbf{Y}_I
\end{bmatrix}\Omega_I, \label{eq:casida}
\end{align}
where $\Omega_I$ is the eigenvalue and the eigenvector $\mathbf{z}_I = \begin{bmatrix}
    \mathbf{X}_I \\ \mathbf{Y}_I
\end{bmatrix}$, with $\mathbf{A}$ and $\mathbf{B}$ defined by
\begin{align}
A_{\mu\nu}&=\omega_\mu\delta_{\mu\nu} + \langle \Phi_\mu^N|\hat{p}^\dagger \hat{r}|\Phi_0^N\rangle v_{pr,qs} \langle \Phi_0^N|\hat{q}^\dagger \hat{s}|\Phi_\nu^N\rangle,\label{eq:mrrpa-A}\\
B_{\mu\nu}&=\langle \Phi_\mu^N|\hat{p}^\dagger \hat{r}|\Phi_0^N\rangle v_{pr,qs} \langle \Phi_\nu^N|\hat{q}^\dagger \hat{s}|\Phi_0^N\rangle.\label{eq:mrrpa-B}
\end{align}
As in standard RPA, the structure of Eq. \eqref{eq:casida} implies that its eigensystem has a paired structure, that is,
the eigenvalues $\Omega_I$ appear in pairs,
\begin{align}
    \Omega_I = - \Omega_{-I},\quad I>0,
\end{align}
where $I>0$ means that $\Omega_I$ is positive, with the corresponding eigenvectors given by
\begin{align}
\mathbf{z}_I = 
\begin{pmatrix}
\mathbf{X}_I \\
\mathbf{Y}_I \\
\end{pmatrix},\quad
\mathbf{z}_{-I} = 
\begin{pmatrix}
\mathbf{Y}_I^* \\
\mathbf{X}_I^* \\   
\end{pmatrix}.\label{eq:casida-pair}
\end{align}

\subsection{Screened interaction $\mathbf{W}(\omega)$} 
Now the correlation part of screened interaction $\mathbf{W}(\omega)$ \eqref{eq:dyson-W} can be evaluated from
\begin{align}
w_{pr,qs}^c(\omega) \equiv& [\mathbf{v}\mathbf{\Pi}(\omega)\mathbf{v}]_{pr,qs} \nonumber\\
=&  \left[\mathbf{v} \mathbf{\Pi}_0(\omega) \mathbf{v}\right]_{pr,qs}
+
 [\mathbf{v}\mathbf{\Pi}_0(\omega) \mathbf{v}
\mathbf{\Pi}_0(\omega) \mathbf{v}]_{pr,qs} + \cdots \nonumber\\
=& \left[\mathbf{v}^L \mathbf{D}_0(\omega) \mathbf{v}^R\right]_{pr,qs} +
    [\mathbf{v}^L\mathbf{D}_0(\omega) \mathbf{V}\mathbf{D}_0(\omega) \mathbf{v}^R]_{pr,qs} + \cdots\nonumber\\
=&
\left[\mathbf{v}^L \mathbf{D}(\omega) \mathbf{v}^R\right]_{pr,qs}\nonumber\\
    =&  \sum_{I>0} \left[\mathbf{v}^L \left(
\frac{\mathbf{z}_I\mathbf{z}_I^{\dagger}}{\omega -\Omega_I +\ii0^+}-\frac{\mathbf{z}_{-I}\mathbf{z}_{-I}^{\dagger}}{\omega + \Omega_I -\ii0^+}\right) \mathbf{v}^R\right]_{pr,qs},
\end{align}
which can be expressed more explicitly by invoking the paired eigenvectors as
\begin{align}
    w_{pr,qs}^c (\omega)
    =&  \sum_{I>0} \left[\mathbf{v}^L \left(
\frac{\mathbf{z}_I\mathbf{z}_I^{\dagger}}{\omega-\Omega_I +\ii0^+}-\frac{\mathbf{z}_{-I}\mathbf{z}_{-I}^{\dagger}}{\omega+\Omega_I -\ii0^+}\right) \mathbf{v}^R\right]_{pr,qs}\nonumber\\
=&  \sum_{I>0} \left[\mathbf{v}^L \left(
\frac{1}{\omega-\Omega_I +\ii0^+}\begin{bmatrix}
    X_{\mu I}X^*_{\nu I} & X_{\mu I}Y_{\nu I}^* \\ Y_{\mu I}X^*_{\nu I} & Y_{\mu I}Y_{\nu I}^*
\end{bmatrix}-\frac{1}{\omega+\Omega_I -\ii0^+}\begin{bmatrix}
    Y^*_{\mu I}Y_{\nu I} & Y^*_{\mu I}X_{\nu I}\\ X^*_{\mu I}Y_{\nu I} & X^*_{\mu I}X_{\nu I}
\end{bmatrix}\right) \mathbf{v}^R\right]_{pr,qs}\nonumber\\
=&  \sum_{I>0}\left(\frac{M_{pr,I}M_{qs,I}^*}{\omega-\Omega_I +\ii0^+}-\frac{M_{pr,I}^*M_{qs,I}}{\omega+\Omega_I-\ii0^+}\right), \label{eq:screen-mr}
\end{align}
with $M_{pr,I}$ defined by
\begin{align}
    M_{pr,I} =& v_{pr,qs}\left(\langle \Phi_0^N|\hq^\dagger\hs|\Phi_\mu^N\rangle X_{\mu I}+\langle \Phi_\mu^N|\hq^\dagger\hs|\Phi_0^N\rangle Y_{\mu I}\right).\label{eq:Mgeneral}
\end{align}

\subsection{Self-energy $\mathbf{\Sigma}^{\text{MR-}GW}(\omega)$} 
The exchange-correlation self-energy diagram (Fig. \ref{fig:framework}b) corresponds to  $\Sigma^{\text{xc},GW} = \ii G_0 W$. The first term in $\mathbf{W}(\omega)$ \eqref{eq:dyson-W} gives the exchange self-energy 
\begin{align}
    \Sigma^{x}_{pq}(\omega) = -v_{pr,sq}\langle \hat{s}^\dagger \hat{r}\rangle_0,
\end{align}
while the second term in $\mathbf{W}(\omega)$ gives the $GW$ correlation self-energy $\Sigma^{c,GW}_{rs}(\omega)$,
which can be evaluated by using the Lehmann representation of the zeroth-order Green's function
\begin{align}
     [G_0]_{pq}(\omega) &= \sum_\mu \frac{\langle\Phi_0|\hp|\Phi_\mu^{N+1}\rangle\langle\Phi_\mu^{N+1}|\hq^\dagger| \Phi_0\rangle}{\omega-\omega_\mu^{N+1}+\ii0^+}+\frac{\langle \Phi_0 |\hq^\dagger|\Phi_\mu^{N-1}\rangle\langle\Phi_\mu^{N-1}|\hp| \Phi_0\rangle}{\omega+\omega_\mu^{N-1}-\ii0^+},
\end{align}
via a convolution in the frequency space as
\begin{align}
    \Sigma^{c,GW}_{pq}(\omega) &=\frac{\ii}{2\pi}\int d\omega^\prime \, [G_0]_{rs}(\omega+\omega^\prime)w^c_{pr,sq}(\omega^\prime)\nonumber\\
    &=\frac{\ii}{2\pi}\sum_{I>0}\sum_{\mu}\int d\omega^\prime \, \left(\frac{\langle \Phi_0^N |\hat{r}|\Phi_\mu^{N+1}\rangle \langle \Phi_\mu^{N+1} |\hat{s}^\dagger|\Phi_0^N\rangle}{\omega^\prime+\omega-\omega_\mu^{N+1}+\ii0^+}+\frac{ \langle \Phi_0^N |\hat{s}^\dagger|\Phi_\mu^{N-1}\rangle \langle \Phi_\mu^{N-1} |\hat{r}|\Phi_0^N\rangle}{\omega^\prime+\omega+\omega_\mu^{N-1}-\ii0^+}\right)\nonumber\\
    &\qquad\qquad\qquad\qquad\qquad\cdot\left(\frac{M_{pr,I}M_{sq,I}^*}{\omega^\prime-\Omega_I +\ii 0^+}-\frac{M_{pr,I}^*M_{sq,I}}{\omega^\prime+\Omega_I-\ii 0^+}\right),\label{eq:sigma-before-int}
\end{align}
The product can be expanded into four terms, two of which after integration vanish, as all the poles lie in the same side of the real axis,
while the other two can be integrated out using contours shown in Fig. \ref{fig:sigma-contour}. 
The 
result reads 
\begin{align}
    \Sigma^{c,GW}_{pq}(\omega) =& \sum_{I>0}\sum_{\mu}
    \left(\frac{\langle \Phi_0^N |\hat{r}|\Phi_\mu^{N+1}\rangle \langle \Phi_\mu^{N+1} |\hat{s}^\dagger|\Phi_0^N\rangle M_{pr,I}^*M_{sq,I}}{\omega-\Omega_I-\omega_\mu^{N+1}+\ii0^+} +  \frac{\langle \Phi_0^N |\hat{s}^\dagger|\Phi_\mu^{N-1}\rangle \langle \Phi_\mu^{N-1} |\hat{r}|\Phi_0^N\rangle M_{pr,I}M_{sq,I}^*}{\omega+\Omega_I+\omega_\mu^{N-1}-\ii0^+}\right).
\end{align}
Finally, adding the first-order self-energy $\mathbf{\Sigma}_1^{11}(\omega)$, the total MR-$GW$ self-energy is 
\begin{align}
    \Sigma_{pq}^{\text{MR-}GW}(\omega) =&
    [\Sigma_1^{11}(\omega)]_{pq}
    +
    \Sigma^{c,GW}_{pq}(\omega) \nonumber\\
    =&
    \, u_{pq}+\bar{v}_{pq,rs}\langle \hat{r}^\dagger\hat{s}\rangle_0 \nonumber\\
    &+\sum_{I>0}\sum_{\mu}\frac{\langle \Phi_0^N |\hat{r}|\Phi_\mu^{N+1}\rangle \langle \Phi_\mu^{N+1} |\hat{s}^\dagger|\Phi_0^N\rangle M_{pr,I}^*M_{sq,I}}{\omega-\Omega_I-\omega_\mu^{N+1}+\ii0^+}\nonumber\\
    &+ \sum_{I>0}\sum_{\mu} \frac{\langle \Phi_0^N |\hat{s}^\dagger|\Phi_\mu^{N-1}\rangle \langle \Phi_\mu^{N-1} |\hat{r}|\Phi_0^N\rangle M_{pr,I}M_{sq,I}^*}{\omega+\Omega_I+\omega_\mu^{N-1}-\ii0^+}, \label{eq:sigma-mrgw-general}
\end{align}
This expression is general in the sense that in the limiting case where $\hH_0$ is quadratic, it reduces to the standard $GW$ self-energy.
Thus, MR-$GW$ can be viewed as a natural generalization of the standard $GW$.

\begin{figure}[!hbt]
    \centering
\begin{tikzpicture}[x=0.75pt,y=0.75pt,yscale=-1,xscale=1, scale=.7]

\draw [color={rgb, 255:red, 0; green, 0; blue, 0 }  ,draw opacity=1 ][line width=0.75]  (112.06,331.04) -- (517.14,331.39)(325.15,144) -- (324.94,384) (510.15,326.39) -- (517.14,331.39) -- (510.14,336.39) (320.14,150.99) -- (325.15,144) -- (330.14,151)  ;
\draw  [color={rgb, 255:red, 208; green, 2; blue, 27 }  ,draw opacity=1 ] (234,335.74) -- (251.14,353)(251.14,335.74) -- (234,353) ;
\draw  [color={rgb, 255:red, 208; green, 2; blue, 27 }  ,draw opacity=1 ] (266.65,335.74) -- (283.79,353)(283.79,335.74) -- (266.65,353) ;
\draw  [color={rgb, 255:red, 208; green, 2; blue, 27 }  ,draw opacity=1 ] (299.3,335.74) -- (316.44,353)(316.44,335.74) -- (299.3,353) ;
\draw  [color={rgb, 255:red, 208; green, 2; blue, 27 }  ,draw opacity=1 ] (331.95,335.74) -- (349.09,353)(349.09,335.74) -- (331.95,353) ;
\draw  [color={rgb, 255:red, 208; green, 2; blue, 27 }  ,draw opacity=1 ] (364.6,335.74) -- (381.74,353)(381.74,335.74) -- (364.6,353) ;
\draw  [color={rgb, 255:red, 208; green, 2; blue, 27 }  ,draw opacity=1 ] (397.24,335.74) -- (414.39,353)(414.39,335.74) -- (397.24,353) ;
\draw  [color={rgb, 255:red, 74; green, 144; blue, 226 }  ,draw opacity=1 ] (127.61,303) -- (144.76,320.26)(144.76,303) -- (127.61,320.26) ;
\draw  [color={rgb, 255:red, 74; green, 144; blue, 226 }  ,draw opacity=1 ] (160.26,303) -- (177.4,320.26)(177.4,303) -- (160.26,320.26) ;
\draw  [color={rgb, 255:red, 74; green, 144; blue, 226 }  ,draw opacity=1 ] (192.91,303) -- (210.05,320.26)(210.05,303) -- (192.91,320.26) ;
\draw  [color={rgb, 255:red, 74; green, 144; blue, 226 }  ,draw opacity=1 ] (225.56,303) -- (242.7,320.26)(242.7,303) -- (225.56,320.26) ;
\draw  [color={rgb, 255:red, 74; green, 144; blue, 226 }  ,draw opacity=1 ] (258.21,303) -- (275.35,320.26)(275.35,303) -- (258.21,320.26) ;
\draw  [color={rgb, 255:red, 74; green, 144; blue, 226 }  ,draw opacity=1 ] (290.86,303) -- (308,320.26)(308,303) -- (290.86,320.26) ;
\draw [color={rgb, 255:red, 189; green, 16; blue, 224 }  ,draw opacity=1 ]   (115,325.24) .. controls (115,132.65) and (511,135.86) .. (515,325.24) ;
\draw [shift={(307.53,182.04)}, rotate = 359.96] [fill={rgb, 255:red, 189; green, 16; blue, 224 }  ,fill opacity=1 ][line width=0.08]  [draw opacity=0] (10.72,-5.15) -- (0,0) -- (10.72,5.15) -- (7.12,0) -- cycle    ;
\draw [color={rgb, 255:red, 189; green, 16; blue, 224 }  ,draw opacity=1 ]   (115,325.24) -- (515,325.24) ;
\draw [shift={(320,325.24)}, rotate = 180] [fill={rgb, 255:red, 189; green, 16; blue, 224 }  ,fill opacity=1 ][line width=0.08]  [draw opacity=0] (10.72,-5.15) -- (0,0) -- (10.72,5.15) -- (7.12,0) -- cycle    ;

\draw (532,312.4) node [anchor=north west][inner sep=0.75pt]  [font=\large]  {$\omega ^{\prime }$};
\draw (338,136.4) node [anchor=north west][inner sep=0.75pt]  [font=\large]  {$\ii \omega ^{\prime }$};
\draw (191,278.4) node [anchor=north west][inner sep=0.75pt]    {$-\Omega _{I} +\ii 0^{+}$};
\draw (332,362.4) node [anchor=north west][inner sep=0.75pt]    {$-\omega +\omega _{\mu }^{N+1} -\ii 0^{+}$};
\end{tikzpicture}
\hfill
\begin{tikzpicture}[x=0.75pt,y=0.75pt,yscale=-1,xscale=1,scale=.7]
\draw [color={rgb, 255:red, 0; green, 0; blue, 0 }  ,draw opacity=1 ][line width=0.75]  (112.06,331.04) -- (517.14,331.39)(325.15,144) -- (324.94,384) (510.15,326.39) -- (517.14,331.39) -- (510.14,336.39) (320.14,150.99) -- (325.15,144) -- (330.14,151)  ;
\draw  [color={rgb, 255:red, 208; green, 2; blue, 27 }  ,draw opacity=1 ] (234,301.74) -- (251.14,319)(251.14,301.74) -- (234,319) ;
\draw  [color={rgb, 255:red, 208; green, 2; blue, 27 }  ,draw opacity=1 ] (266.65,301.74) -- (283.79,319)(283.79,301.74) -- (266.65,319) ;
\draw  [color={rgb, 255:red, 208; green, 2; blue, 27 }  ,draw opacity=1 ] (299.3,301.74) -- (316.44,319)(316.44,301.74) -- (299.3,319) ;
\draw  [color={rgb, 255:red, 208; green, 2; blue, 27 }  ,draw opacity=1 ] (331.95,301.74) -- (349.09,319)(349.09,301.74) -- (331.95,319) ;
\draw  [color={rgb, 255:red, 208; green, 2; blue, 27 }  ,draw opacity=1 ] (364.6,301.74) -- (381.74,319)(381.74,301.74) -- (364.6,319) ;
\draw  [color={rgb, 255:red, 208; green, 2; blue, 27 }  ,draw opacity=1 ] (397.24,301.74) -- (414.39,319)(414.39,301.74) -- (397.24,319) ;
\draw  [color={rgb, 255:red, 74; green, 144; blue, 226 }  ,draw opacity=1 ] (329.61,335) -- (346.76,352.26)(346.76,335) -- (329.61,352.26) ;
\draw  [color={rgb, 255:red, 74; green, 144; blue, 226 }  ,draw opacity=1 ] (362.26,335) -- (379.4,352.26)(379.4,335) -- (362.26,352.26) ;
\draw  [color={rgb, 255:red, 74; green, 144; blue, 226 }  ,draw opacity=1 ] (394.91,335) -- (412.05,352.26)(412.05,335) -- (394.91,352.26) ;
\draw  [color={rgb, 255:red, 74; green, 144; blue, 226 }  ,draw opacity=1 ] (427.56,335) -- (444.7,352.26)(444.7,335) -- (427.56,352.26) ;
\draw  [color={rgb, 255:red, 74; green, 144; blue, 226 }  ,draw opacity=1 ] (460.21,335) -- (477.35,352.26)(477.35,335) -- (460.21,352.26) ;
\draw  [color={rgb, 255:red, 74; green, 144; blue, 226 }  ,draw opacity=1 ] (492.86,335) -- (510,352.26)(510,335) -- (492.86,352.26) ;
\draw [color={rgb, 255:red, 189; green, 16; blue, 224 }  ,draw opacity=1 ]   (115,325.24) .. controls (115,132.65) and (511,135.86) .. (515,325.24) ;
\draw [shift={(307.53,182.04)}, rotate = 359.96] [fill={rgb, 255:red, 189; green, 16; blue, 224 }  ,fill opacity=1 ][line width=0.08]  [draw opacity=0] (10.72,-5.15) -- (0,0) -- (10.72,5.15) -- (7.12,0) -- cycle    ;
\draw [color={rgb, 255:red, 189; green, 16; blue, 224 }  ,draw opacity=1 ]   (115,325.24) -- (515,325.24) ;
\draw [shift={(320,325.24)}, rotate = 180] [fill={rgb, 255:red, 189; green, 16; blue, 224 }  ,fill opacity=1 ][line width=0.08]  [draw opacity=0] (10.72,-5.15) -- (0,0) -- (10.72,5.15) -- (7.12,0) -- cycle    ;
\draw (532,312.4) node [anchor=north west][inner sep=0.75pt]  [font=\large]  {$\omega ^{\prime }$};
\draw (338,136.4) node [anchor=north west][inner sep=0.75pt]  [font=\large]  {$\ii\omega ^{\prime }$};
\draw (371,358.4) node [anchor=north west][inner sep=0.75pt]    {$\Omega _{I} -\ii 0^{+}$};
\draw (201-20,272.4) node [anchor=north west][inner sep=0.75pt]    {$-\omega -\omega _{\mu }^{N-1} +\ii 0^{+}$};
\end{tikzpicture}
        \caption{Integration contours for Eq. \eqref{eq:sigma-before-int}.}
    \label{fig:sigma-contour}
\end{figure}

\subsection{Simplifications for $\hat{H}_0^{\mathrm{Dyall}}$}
The four kinds of screening effects [Fig. \ref{fig:framework}(d)] lead to a 4-by-4 block structure\cite{wang_generalized_2025} of $\mathbf{A}$ and $\mathbf{B}$ in Eqs. \eqref{eq:mrrpa-A} and \eqref{eq:mrrpa-B},
\begin{align}
\mathbf{A} =
\begin{pmatrix}
[A_{ai,bj}] & [A_{ai,\nu j}] & [A_{ai,b\nu}] & [A_{ai,\nu}] \\
[A_{\mu i,bj}] &[A_{\mu i,\nu j}] & [A_{\mu i,b\nu}] & [A_{\mu i,\nu}] \\
[A_{a\mu,bj}] & [A_{a\mu,\nu j}] & [A_{a\mu,b\nu}] & [A_{a\mu,\nu}] \\
[A_{\mu,bj}] &  [A_{\mu,\nu j}] & [A_{\mu,b\nu}] & [A_{\mu,\nu}]\\
\end{pmatrix},\quad
\mathbf{B} =
\begin{pmatrix}
[B_{ai,bj}] & [B_{ai,\nu j}] & [B_{ai,b\nu}] & [B_{ai,\nu}] \\
[B_{\mu i,bj}] &[B_{\mu i,\nu j}] & [B_{\mu i,b\nu}] & [B_{\mu i,\nu}] \\
[B_{a\mu,bj}] & [B_{a\mu,\nu j}] & [B_{a\mu,b\nu}] & [B_{a\mu,\nu}] \\
[B_{\mu,bj}] &  [B_{\mu,\nu j}] & [B_{\mu,b\nu}] & [B_{\mu,\nu}]\\
\end{pmatrix},\label{eq:AandB4by4}
\end{align}
where indices $\mu$ and $\nu$ now refer to eigenstates in the active space,
see Eq. \eqref{eq:active-eigen}, while other indices $\{i,j,k,\cdots\}$ and $\{a,b,c,\cdots\}$ refer to core and virtual orbitals, respectively.
The detailed expressions of these matrix elements can be found in Ref. \cite{wang_generalized_2025}. Throughout this work, we assume that the total Hamiltonian is real and hence all quantities are real. Consequently, $M_{pr,I}$ in Eq. \eqref{eq:Mgeneral} simplifies as
\begin{align}
 M_{pr,I} =& v_{pr,qs}\langle \Phi_\mu^N|\hq^\dagger\hs|\Phi_0^N\rangle R_{\mu I},\quad R_{\mu I} \equiv X_{\mu I}+ Y_{\mu I}.\label{eq:W-M-general}
\end{align}
Following Eq. \eqref{eq:AandB4by4}, the matrix $\mathbf{R}$ has a block structure
\begin{align}
    \mathbf{R} = \begin{pmatrix}
        [R_{ai,I}] \\ [R_{\mu i,I}] \\ [R_{a \mu, I}] \\ [R_{\mu,I}]
    \end{pmatrix},
\end{align}
such that Eq. \eqref{eq:W-M-general} can be written explicitly as
\begin{align}
    M_{pr,I} =&(pr|ai) R_{ai,I} + (pr|\mu i) R_{\mu i,I} + (pr|a \mu) R_{a \mu, I}+(pr|\mu) R_{\mu, I},
\end{align}
where the `dressed' integrals are defined by
\begin{align}
    (pr|\mu i) \equiv& (-1)(pr|xi)D^{[+1]}_{\mu,x}, \\
    (pr|a\mu)  \equiv& (pr|ax)D^{[-1]}_{\mu,x}, \\
    (pr|\mu)    \equiv& \left(1-\delta_{p\in A}\delta_{r\in A}\right)(pr|xy)D^{[0]}_{\mu,xy},
\end{align}
with the transition density matrices for adding, removing and exciting active electrons given by
\begin{align}
    D_{\mu, x}^{[+1]} &= \langle \Xi^A_{\mu,N_a+1}|\hat{x}^\dagger|\Xi^A_{0,N_a}\rangle,\label{eq:tdm-cas-1}\\
    D_{\mu, x}^{[-1]} &= \langle \Xi^A_{\mu,N_a-1}|\hat{x}|\Xi^A_{0,N_a}\rangle,\label{eq:tdm-cas-2}\\
    D_{\mu, xy}^{[0]} &= \langle \Xi^A_{\mu,N_a}|\hat{x}^\dagger\hat{y}|\Xi^A_{0,N_a}\rangle.\label{eq:tdm-cas-3}
\end{align}
The final MR-$GW$ self-energy, which constitutes the central result of this work, is obtained from Eq. \eqref{eq:sigma-mrgw-general} as
\begin{align}
    \Sigma_{pq}^{\text{MR-}GW}(\omega) =& u_{pq}+\bar{v}_{pq,ii} + \bar{v}_{pq,xy}\langle \hat{x}^\dagger\hat{y}\rangle_0 \nonumber\\
    &+\sum_{I>0}\frac{M_{ap,I}M_{aq,I}}{\omega-\Omega_I-\epsilon_a+\ii 0^+}+\sum_{I>0}\sum_{\mu}\frac{D_{\mu,x}^{[+1]}D_{\mu,y}^{[+1]}M_{xp,I}M_{yq,I}}{\omega-\Omega_I-\omega_{\mu,N_a+1}^A+\ii 0^+}\nonumber\\
    &+ \sum_{I>0} \frac{M_{pi,I}M_{qi,I}}{\omega+\Omega_I-\epsilon_i-\ii 0^+} + \sum_{I>0}\sum_{\mu} \frac{D_{\mu,x}^{[-1]}D^{[-1]}_{\mu,y}M_{px,I}M_{qy,I}}{\omega+\Omega_I+\omega_{\mu,N_a-1}^A-\ii 0^+}.\label{eq:se-mrgw-cas}
\end{align}
With the self-energy, the Green's function is obtained from Eq. \eqref{eq:dyson-mrgw}, viz.,
\begin{align}
    \mathbf{G}(\omega) = \left([\mathbf{G}_0(\omega)]^{-1} - \boldsymbol{\Sigma}^{\text{MR-}GW}(\omega)\right)^{-1},\label{eq:G-final}
\end{align}
which is used to compute the spectral function $A(\omega)$.

{\color{\mycolor}
\section{Workflow and computational scaling}
Appendix D of the main text summarizes the computational workflow and
provides a direct cost comparison with standard $GW$. Here, we describe
the present implementation in detail and outline a prospective
frequency-domain response formulation that avoids full spectral
diagonalization of the active-space Hamiltonian and explicit solution
of the dense MR-RPA eigenvalue problem.
}

{\color{\mycolor}
\subsection{Detailed workflow and scaling}
We use $N_{\rm orb}=N_c+N_a+N_v$, $D$ as the largest determinant
dimension among the three active particle-number sectors,
$K\leq N_cN_v+D(N_c+N_v+1)$ as the dimension of each MR-RPA
$\mathbf A$ or $\mathbf B$ block, and $L\leq N_c+N_v+2D=\mathcal{O}(N_c+N_v+D)$
as the number of charged active-space or
inactive-orbital Green's-function channels entering
Eq.~\eqref{eq:se-mrgw-cas}. 
We also use $N_\omega$
frequency points and $N_\Sigma=\mathcal{O}(N_{\rm orb})$ or
$\mathcal{O}(N_{\rm orb}^2)$ for diagonal or full self-energy matrices,
respectively. Numerical constants and lower-order terms are omitted
below.

\textbf{Step 1: Integral transformation.---}A conventional four-index
AO--MO transformation requires $\mathcal{O}(N_{\rm orb}^5)$ work. This step can
be avoided or reduced by density fitting/RI, as discussed below.

\textbf{Step 2: Active-sector solutions.---}Diagonalizing the Dyall Hamiltonian in the
neutral, electron-attached, and electron-removed sectors
[Eq.~\eqref{eq:Hdyall-main2}] scales as $\mathcal{O}(D^3)$.

\textbf{Step 3: Transition density matrices.---}Transforming the active-space
one-body and particle-addition/removal transition density matrices in
Eqs.~\eqref{eq:tdm-cas-1}--\eqref{eq:tdm-cas-3} requires
$\mathcal{O}[D^2(N_a^2+2N_a)]=\mathcal{O}(D^2N_a^2)$ work when all dense sector
eigenvectors are retained.

\textbf{Step 4: MR-RPA solution.---}Constructing and diagonalizing the
dense MR-RPA problem in Eqs.~\eqref{eq:AandB4by4} and
\eqref{eq:MRRPAdiag} scales as $\mathcal{O}(K^3)$.

\textbf{Step 5: Screened-interaction couplings.---}Constructing all
$M_{pr,I}$ from Eqs.~\eqref{eq:Mgeneral} and \eqref{eq:W-M-general} requires
$\mathcal{O}(N_{\rm orb}^2K^2)$ work. 

\textbf{Step 6: Self-energy.---}After precontracting the
active-space transition densities with $M_{xp,I}$ or $M_{px,I}$,
Eq.~\eqref{eq:se-mrgw-cas} contains $\mathcal{O}(K)$ neutral
MR-RPA modes and $\mathcal{O}(L)$ Green's-function channels for each
requested orbital pair. Its evaluation at $N_\omega$ frequencies
therefore requires $\mathcal{O}(N_\omega N_\Sigma K L)$ work. The
precontractions themselves require at most
$\mathcal{O}(K N_{\rm orb}N_aD)$ work and are performed once before the
frequency loop.

\textbf{Step 7: Dyson equation.---}Solving Eq.~\eqref{eq:G-final} by
full matrix inversion at every frequency requires
$\mathcal{O}(N_\omega N_{\rm orb}^3)$ work. 

For $N_a=0$, $K=\mathcal{O}(N_cN_v)=\mathcal{O}(N_{\rm orb}^2)$, and Steps 4 and 5 reduce to
the corresponding full-pole dense spectral $G_0W_0$ operations with
$\mathcal{O}(N_{\rm orb}^6)$ cost.\cite{van_setten_gw100_2015} RI-based
frequency-domain and space--time $G_0W_0$ implementations avoid these
dense operations and can reduce selected-quasiparticle calculations to
$\mathcal{O}(N_{\rm orb}^4)$ and $\mathcal{O}(N_{\rm orb}^3)$, respectively.
\cite{ren_resolution--identity_2012,govoni_large_2015,
wilhelm_low-scaling_2021,zhu_all-electron_2021}

\subsection{Frequency-domain response formulation}
The explicit spectral implementation requires all relevant
active-sector eigenstates. An alternative is to evaluate the
frequency-dependent response as the action of a resolvent, following
the direct iterative strategies used in FCI and DMRG response
theory.\cite{olsen_large_response_1988,
koch_fci_response_1991,dorando_analytic_2009} Closely related
frequency-dependent linear-equation strategies also underlie
Sternheimer formulations of RPA and $GW$.\cite{govoni_large_2015,
peng_basis-set-error-free_2023}
Starting from Eq. \eqref{eq:mr-rpa-pi0}, we rewrite $\Pi_0$ as
\begin{align}
    [\Pi_0]_{pr,qs}(\omega)  =& \langle \Phi_0|\hp^\dagger \hr |X_{qs}(\omega)\rangle + \langle \Phi_0|\hq^\dagger \hs |X_{pr}(-\omega)\rangle,\label{eq:X-noRI}
\end{align}
Introducing $\hat{\mathcal Q}=1-|\Phi_0\rangle\langle\Phi_0|$, the required
response vector is the projected resolvent action
\begin{align}
|X_{pr}(\omega)\rangle \equiv
\sum_{\mu>0}
\frac{|\Phi_\mu\rangle
\langle\Phi_\mu|\hp^\dagger\hr|\Phi_0\rangle}
{\omega-\omega_\mu+\ii 0^+}
=
\hat{\mathcal Q}\,[\omega-(\hH_0-E_0)+\ii 0^+]^{-1}
\hat{\mathcal Q}\hp^\dagger\hr|\Phi_0\rangle,\label{eq:response-vector}
\end{align}
which can be
evaluated by solving a linear system at each frequency, without an
explicit sum over all excited states.

Given that the zeroth-order ground state is a CASCI/CASSCF wavefunction
$|\Phi^I_0\rangle|\Xi_{0,N_a}^A\rangle$ with 
$|\Xi^A_{0,N_a}\rangle = \sum_\lambda |\Phi^A_{\lambda,N_a}\rangle C^A_{\lambda,N_a}$, 
$|X_{pr}(\omega)\rangle$ contains four non-vanishing blocks, 
corresponding to four kinds of excitations:

1. $|X_{ai}(\omega)\rangle$ from core orbitals to virtual:
\begin{align}
    |X_{ai}(\omega)\rangle=& \frac{|\Phi_i^a\rangle|\Xi_{0,N_a}^A\rangle}{\omega-(\epsilon_a-\epsilon_i)},
\end{align}
where we use the notation $|\Phi_i^a\rangle = \hat{a}^\dagger \hat{i}|\Phi_0^I\rangle$ and
the infinitesimal frequency shifts are suppressed.

2. $|X_{xi}(\omega)\rangle$ from core orbitals to active:
\begin{align}
|X_{xi}(\omega)\rangle = |\Phi_i\rangle \sum_{\lambda}|\Phi^A_{\lambda,{N_a+1}}\rangle C^{[+1]}_{\lambda,xi}(\omega),
\end{align}
where $|\Phi_i\rangle = \hat{i}|\Phi^I\rangle$, and the expansion coefficient $C^{[+1]}_{\lambda,xi}(\omega)$ is determined by the following response equation
\begin{align}
\sum_{\lambda}\left[\delta_{\nu\lambda}(\omega+E_{0,N_a}^A+\epsilon_i)-\langle\Phi^A_{\nu,N_a+1}|\hH_0^\text{act}|\Phi^A_{\lambda,N_a+1}\rangle\right] C^{[+1]}_{\lambda,xi}(\omega)=(-1)\langle\Phi^A_{\nu,N_a+1}|\hx^\dagger|\Xi_{0,N_a}^A\rangle.
\end{align}

3. $|X_{ax}(\omega)\rangle$ from active orbitals to virtual:
\begin{align}
|X_{ax}(\omega)\rangle = |\Phi^a\rangle \sum_{\lambda}|\Phi^A_{\lambda,N_a-1}\rangle 
C^{[-1]}_{\lambda,ax}(\omega),
\end{align}
where $|\Phi^a\rangle = \hat{a}^\dagger|\Phi^I\rangle$, and the expansion coefficient $C^{[-1]}_{\lambda,ax}(\omega)$ is determined by
\begin{align}
\sum_{\lambda}\left[\delta_{\nu\lambda}(\omega+E_{0,N_a}^A-\epsilon_a)-\langle\Phi^A_{\nu,N_a-1}|\hH_0^\text{act}|\Phi^A_{\lambda,N_a-1}\rangle\right] C^{[-1]}_{\lambda,ax}(\omega)=\langle\Phi^A_{\nu,N_a-1}|\hx|\Xi_{0,N_a}^A\rangle.
\end{align}

4. $|X_{xy}(\omega)\rangle$ from active orbitals to active:
\begin{align}
|X_{xy}(\omega)\rangle = |\Phi^I_0\rangle
\sum_{\lambda}|\Phi^A_{\lambda,N_a}\rangle C^{[0]}_{\lambda,xy}(\omega),
\end{align}
with $C^{[0]}_{\lambda,xy}(\omega)$ determined by
\begin{align}
\sum_{\lambda}\left[\delta_{\nu\lambda}(\omega+E_{0,N_a}^A)-\langle\Phi^A_{\nu,N_a}|\hH_0^\text{act}|\Phi^A_{\lambda,N_a}\rangle\right] C^{[0]}_{\lambda,xy}(\omega) = \langle\Phi^A_{\nu,N_a}|\hx^\dagger\hy|\Xi^A_{0,N_a}\rangle - C^A_{\nu,N_a}\langle\hat{x}^\dagger\hat{y}\rangle_0.
\end{align}

The Hamiltonian matrices on the left-hand sides,
$\langle\Phi^A_{\nu,N_a\pm 1}|\hH_0^\text{act}|\Phi^A_{\lambda,N_a\pm 1}\rangle$ 
and
$\langle\Phi^A_{\nu,N_a}|\hH_0^\text{act}|\Phi^A_{\lambda,N_a}\rangle$,
are sparse due to the Slater--Condon rules. This sparsity
enables matrix-vector products and iterative linear solves without
constructing dense Hamiltonian matrices.\cite{olsen_large_response_1988,koch_fci_response_1991}
Let $N_{\rm it}$ be an upper bound on the
iteration count over the three particle-number sectors and all
frequencies. For a two-body active-space Hamiltonian, the
Slater--Condon rules connect each determinant to at most $\mathcal{O}(N_a^4)$
others, so the number of nonzero elements in any sector is bounded by
$\mathcal{O}(DN_a^4)$. The electron-addition, electron-removal, and neutral
responses contain $\mathcal{O}(N_aN_c)$, $\mathcal{O}(N_aN_v)$, and $\mathcal{O}(N_a^2)$ right-hand sides, respectively. Solving all of them at $N_\omega$ frequencies
therefore scales as
$\mathcal{O}\!\left[N_\omega N_{\rm it}(DN_a^4)
\left(N_aN_c+N_aN_v+N_a^2\right)\right]
=\mathcal{O}(N_\omega N_{\rm it}DN_a^5N_{\rm orb})$.

With $C^{[+1]}_{\lambda,xi}(\omega)$, $C^{[-1]}_{\lambda,ax}(\omega)$ and $C^{[0]}_{\lambda,xy}(\omega)$ solved, we reach the zeroth-order irreducible polarizability as
\begin{align}
\boldsymbol{\Pi}_0(\omega)= \begin{bmatrix}
    \mathbf{0}              & [[\Pi_0]_{ia,bj}(\omega)] & \mathbf{0} & \mathbf{0} & \mathbf{0} & \mathbf{0} & \mathbf{0} \\
    [[\Pi_0]_{ai,jb}(\omega)]   & \mathbf{0}            & \mathbf{0} & \mathbf{0} & \mathbf{0} & \mathbf{0} & \mathbf{0} \\
    \mathbf{0}              & \mathbf{0}            & \mathbf{0} & [[\Pi_0]_{ix,yj}(\omega)] & \mathbf{0} & \mathbf{0} & \mathbf{0}\\
    \mathbf{0}              & \mathbf{0}            & [[\Pi_0]_{xi,jy}(\omega)] & \mathbf{0} & \mathbf{0} & \mathbf{0} & \mathbf{0}\\
    \mathbf{0}              & \mathbf{0}            & \mathbf{0} & \mathbf{0} & \mathbf{0} & [[\Pi_0]_{xa,by}(\omega)] & \mathbf{0} \\
    \mathbf{0}              & \mathbf{0}            & \mathbf{0} & \mathbf{0} & [[\Pi_0]_{ax,yb}(\omega)] & \mathbf{0} & \mathbf{0} \\
    \mathbf{0}              & \mathbf{0}            & \mathbf{0} & \mathbf{0} & \mathbf{0} & \mathbf{0} & [[\Pi_0]_{xy,zw}(\omega)]
    \end{bmatrix},
\end{align}
whose building blocks are 
\begin{align}
    [\Pi_0]_{ia,bj}(\omega) &= \frac{\delta_{ij}\delta_{ab}}{\omega-(\epsilon_a-\epsilon_i)},\nonumber\\
    [\Pi_0]_{ai,jb}(\omega) &= \frac{-\delta_{ij}\delta_{ab}}{\omega+(\epsilon_a-\epsilon_i)},\\
    [\Pi_0]_{ix,yj}(\omega) &= \sum_{\lambda}(-1)\delta_{ij}\langle\Xi^A_{0,N_a}|\hx|\Phi^A_{\lambda,N_a+1}\rangle C^{[+1]}_{\lambda,yj}(\omega),\nonumber\\
    [\Pi_0]_{xi,jy}(\omega) &= \sum_{\lambda}(-1)\delta_{ij}\langle \Xi^A_{0,N_a}|\hy|\Phi^A_{\lambda,N_a+1}\rangle C^{[+1]}_{\lambda,xi}(-\omega),\\
    [\Pi_0]_{xa,by}(\omega) &= \sum_{\lambda}\delta_{ab}\langle \Xi^A_{0,N_a}|\hx^\dagger|\Phi^A_{\lambda,N_a-1}\rangle C^{[-1]}_{\lambda,by}(\omega),\nonumber\\
    [\Pi_0]_{ax,yb}(\omega) &= \sum_{\lambda}\delta_{ab}\langle \Xi^A_{0,N_a}|\hy^\dagger|\Phi^A_{\lambda,N_a-1}\rangle C^{[-1]}_{\lambda,ax}(-\omega),\\
  [\Pi_0]_{xy,zw}(\omega) &= 
\sum_{\lambda}
\big[  
\langle \Xi^A_{0,N_a}|\hx^\dagger\hy|\Phi^A_{\lambda,N_a}\rangle C^{[0]}_{\lambda,zw}(\omega) + \langle \Xi^A_{0,N_a}|\hz^\dagger  \hat{w}|\Phi^A_{\lambda,N_a}\rangle C^{[0]}_{\lambda,xy}(-\omega)\big].
\end{align}
The dynamical MR-$GW$ self-energy can then be assembled as in conventional
$GW$,\cite{van_setten_gw100_2015,ren_resolution--identity_2012,
govoni_large_2015,zhu_all-electron_2021}
\begin{align}
\mathbf{\Pi}(\omega)
&
=\mathbf{\Pi}_0(\omega)
[\mathbf{I}-\mathbf{v}\mathbf{\Pi}_0(\omega)]^{-1}, \nonumber\\
\mathbf{w}^c(\omega)
&
=\mathbf{v}\mathbf{\Pi}(\omega)\mathbf{v},\nonumber\\
\Sigma_{rs}^{c,GW}(\omega)
&
=\frac{\ii}{2\pi}
\int_{-\infty}^{\infty}d\omega^\prime\,
\sum_{tu}
[G_0]_{tu}(\omega+\omega^\prime)
w^c_{rt,us}(\omega^\prime).
\label{eq:response-sigma}
\end{align}

The main benefit of the response formulation is
therefore to avoid the explicit sum over active-sector eigenstates,
the dense $\mathcal{O}(K^3)$ MR-RPA diagonalization, and the associated
$\mathcal{O}(K^2)$ eigenvector storage, which are replaced by
sparse, frequency-dependent linear
solves. However, it does not, by itself, reduce the cost of four-index Coulomb
contractions in the orbital space. A practical low-scaling MR-$GW$ implementation should consequently combine the active-space response with the same RI and
frequency-domain techniques used in modern $G_0W_0$.\cite{ren_resolution--identity_2012,govoni_large_2015,
wilhelm_low-scaling_2021,zhu_all-electron_2021}
For selected self-energy or quasiparticle matrix elements, a schematic cost estimate for
a prospective RI implementation is
$\mathcal{O}\!\left(N_{\rm orb}^4
+N_\omega N_{\rm it}DN_a^5N_{\rm orb}\right)$,
where the first term is the conventional RI-$G_0W_0$ orbital-space
work and the second is the active-space response. 
Thus, for a fixed active space (fixed $N_a$ and $D$), this estimate has the same asymptotic orbital-space scaling as the selected modern $G_0W_0$ implementation. 
For larger active spaces, DMRG-based response
solvers can alleviate the exponential determinant-space bottleneck by
representing the response vectors as matrix-product states rather than in the full determinant basis.\cite{dorando_analytic_2009}
}

\section{Computational details and additional results}
\subsection{Computational details}
We implemented MR-$GW$ in the \textsc{MRMBPT}
code\cite{MRMBPTcode}, with molecular integrals and transition density matrices generated from PySCF. Standard $GW$ calculations based on RHF or UHF references were also performed for comparison.
The equilibrium geometry of \ce{O3} was taken from the experimental value listed in the CCCBDB database.\cite{cccbdb_nist}
The HF, CASCI/CASSCF, FCI, and non-Dyson ADC(2) and ADC(3) calculations were performed using the PySCF package.\cite{sun_recent_2020}
{\color{\mycolor}
EOM-CCSD and EOM-CC2 calculations were performed using the Q-Chem package.\cite{epifanovsky_software_2021}
Dyson ADC(2) and ADC(3) calculations were performed using the QuAcK package.\cite{QuAcK}
DMRG calculations were performed using the Block2 package.\cite{zhai_block2_2023}}

{\color{\mycolor}
\subsection{Additional results for \ce{Be}
with different methods and basis sets}
Table \ref{tab:basisset-be} summarizes
the results obtained
with different methods and basis sets for
the principal and satellite peaks.
For closed-shell Be, allowing an unrestricted HF reference changes the $GW$@RHF values by at most 0.25 eV over the three basis sets and leaves the satellite far above the correlated reference value. Spin-symmetry relaxation therefore does not resolve the satellite failure in this case.
}


\begin{table}[!hbt]
    \centering
        \renewcommand{\arraystretch}{1.3}
    \setlength{\tabcolsep}{.2cm}
    \caption{Results for the $(2s)^{-1}$ ionization and satellite energies (in eV) of the \ce{Be} atom calculated with different methods and basis sets. 
    }
    \begin{tabular}{c|cc|cc|cc}
    \hline\hline
            &  \multicolumn{2}{c|}{6-31G} & \multicolumn{2}{c|}{cc-pVDZ} & \multicolumn{2}{c}{cc-pVTZ} \\
    \hline
    Methods          & $(2s)^{-1}$ & satellite & $(2s)^{-1}$ & satellite & $(2s)^{-1}$ & satellite \\
    \hline
    $GW$@RHF         & 8.73 & 21.57 & 8.99 & 20.78 & 9.06 & 19.94 \\
    $GW$@UHF         & 8.74 & 21.82 & 9.00 & 20.93 & 9.08 & 20.13 \\
    CAS(2,4) & 9.38 & 13.25 & 9.51 & 13.32 & 9.52 & 13.38\\
    MR-$GW$(2,4) & 9.27 & 13.27 & 9.53 & 13.35 & 9.61 & 13.42 \\
    Non-Dyson ADC(2) & 8.64 & 18.64 & 8.86 & 18.40 & 8.90 & 18.19\\
    Non-Dyson ADC(3) & 8.88 & 11.93 & 9.04 & 12.06 & 9.04 & 12.02\\
    Dyson ADC(2)     & 8.63 & 18.80 & 8.85 & 18.54 & 8.89 & 18.31 \\
    Dyson ADC(3)     & 9.27 & 12.35 & 9.41 & 12.44 & 9.41 & 12.39 \\
    EOM-CC2 & 8.62 & /$^a$ & 8.84 & /$^a$ & 8.88 & /$^a$ \\
    EOM-CCSD & 9.18 & 13.19 & 9.29 & 13.28 & 9.29 & 13.30 \\
    FCI     & 9.18 & 13.19 & 9.29 & 13.27 & 9.29 & 13.25\\
    \hline\hline
    \end{tabular}\\
    $^a$ EOM-CC2 does not yield an identifiable satellite.
    \label{tab:basisset-be}
\end{table}



\subsection{Additional results for \ce{H2}}
\subsubsection{Self-energies for stretched \ce{H2}}

\begin{figure}[!hbt]
    \centering
    \includegraphics[width=\linewidth]{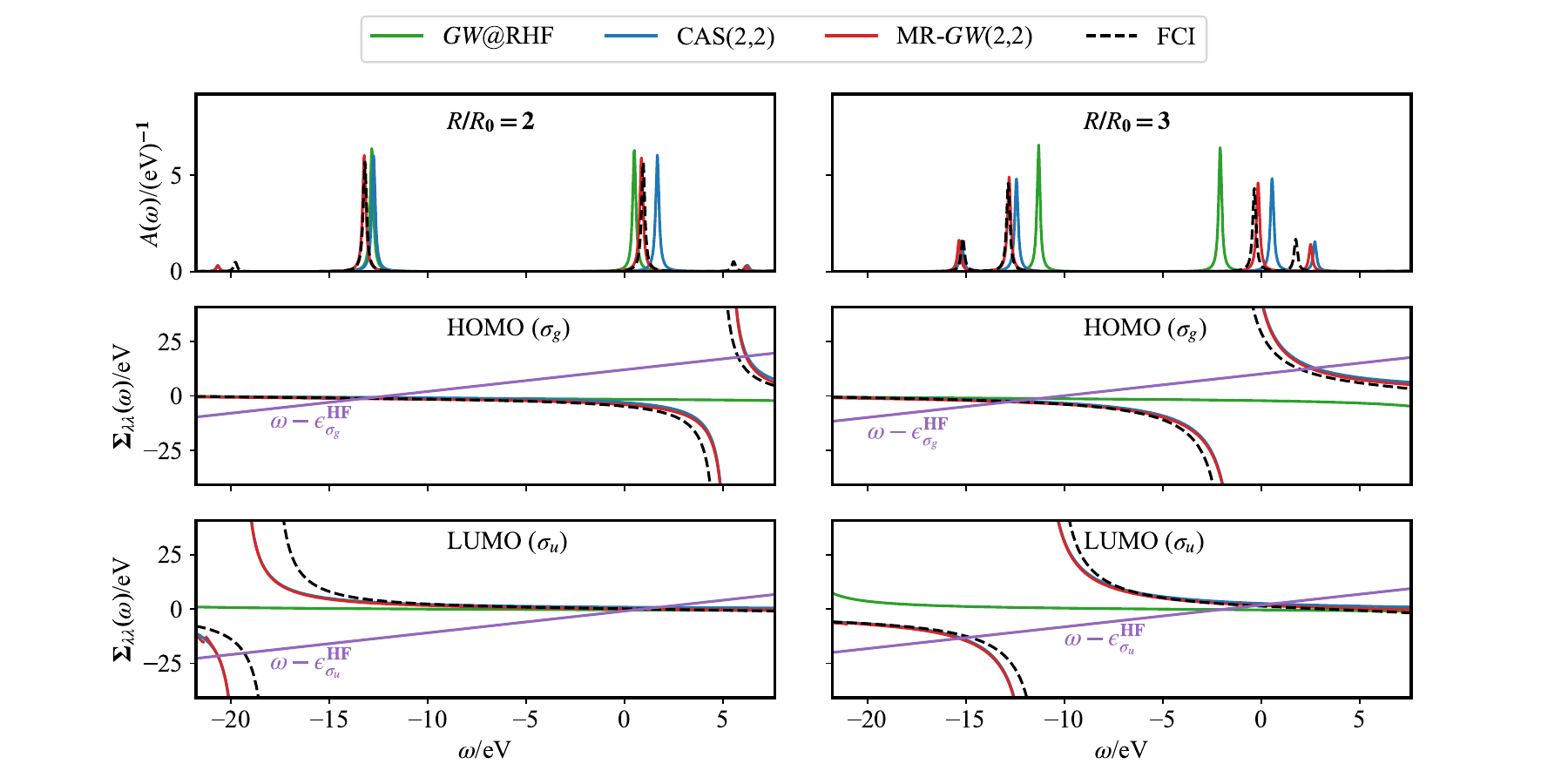}
    \caption{Self-energies on the right-hand side of Eq. \eqref{eq:dyson-diag} and graphical solutions of Eq. \eqref{eq:dyson-diag} (purple) for the HOMO ($\sigma_g$) and LUMO ($\sigma_u$) orbitals of 
    stretched \ce{H2}, where the bond distances $R$ are twice (left) and three times (right) the equilibrium bond length $R_0$, respectively.
    The HOMO (LUMO) orbital energies at $R/R_0=2$ and $R/R_0=3$ are \num{-0.443} (0.035) and \num{-0.371} (\num{-0.065}) Hartrees, respectively.
    }
\end{figure}

\clearpage
{\color{\mycolor}
\subsubsection{Starting-orbital dependence for IP and EA}
Because the present MR-$GW$ approximation is evaluated in a one-shot
manner, its numerical results retain some dependence on the starting
orbitals, analogous to the familiar starting-point dependence of
$G_0W_0$. To assess its magnitude, Table~\ref{tab:orb-CASvsRHF}
compares IPs and EAs obtained using CASSCF and RHF orbitals.
The orbital choice does not alter the qualitative spectral assignments,
and for both orbital choices, the inclusion of the interacting
active-space reference leads to the same qualitative improvement over
single-reference $GW$ at stretched geometries. 
Specifically, the principal IPs are only weakly affected by the choice of orbitals,
whereas the EAs exhibit a more noticeable dependence, particularly
for the higher-energy satellite. 
For the EA benchmark considered here, MR-$GW$
based on RHF orbitals happens to agree more closely with FCI than the
corresponding calculation based on CASSCF orbitals. 
It may reflect the different orbital-relaxation requirements
of the neutral and electron-attached states: the CASSCF $\sigma_u$ orbital is optimized for the neutral reference, whereas the RHF $\sigma_u$
orbital provides a more favorable starting
description of the electron-attached state.
Introducing partial self-consistency or optimizing the
orbitals with information from the charged sectors may provide a
route toward reducing this dependence in future work.}

\begin{table}[!hbt]
    \centering
    \renewcommand{\arraystretch}{1.3}
    \setlength{\tabcolsep}{.2cm}
    \caption{Comparison of MR-$GW$ with CASSCF and RHF orbitals for the prediction of IP and EA (in eV) for \ce{H2} in the cc-pVTZ basis set.
    $GW$@RHF and FCI results are also presented as reference.
    }
    \begin{tabular}{cc|cc|ccc|c}
    \hline\hline
    &                  & \multicolumn{2}{c|}{CASSCF orbitals} & \multicolumn{3}{c|}{RHF orbitals} & \\
    \cline{3-7}
    &                  & CAS(2,2)           & MR-$GW$(2,2)   & $GW$@RHF &  CAS(2,2)  & MR-$GW$(2,2) & FCI \\
    \hline
    \multirow{2}{*}{$R/R_0=1$} & IP: $(\sigma_g)^{-1}$ &  16.62   & 16.60  &  16.48   & 16.21 &  16.50 &  16.40 \\
                               & EA: $(\sigma_u)^{+1}$ & \num{-17.46}  & \num{-4.50} & \num{-4.28}  &   \num{-4.59}    &  \num{-4.26}     &  \num{-4.21} \\
    \hline
    \multirow{4}{*}{$R/R_0=2$} & IP: $(\sigma_g)^{-1}$ &  13.35  & 13.64 & 12.84 & 12.75  &  13.22 & 13.19\\
                               & IP: satellite         &  19.60  & 19.74 & / & 20.67  &  20.66    & 19.77\\
                              & EA: $(\sigma_u)^{+1}$ & \num{-4.84}  & \num{-1.88} & \num{-0.50}& \num{-1.66}  & \num{-0.86}   & \num{-0.94}\\
                              & EA: satellite & \num{-12.15}  & \num{-9.24}   &   /     & \num{-6.23}  &   \num{-6.16}   & \num{-5.53}\\
    \hline
    \multirow{4}{*}{$R/R_0=3$} & IP: $(\sigma_g)^{-1}$     & 12.96 & 13.38  & 11.30 & 12.43 & 12.80 & 12.85\\
                                         & IP: satellite   & 15.12 & 15.36  & / & 15.21 & 15.34 & 15.15\\
                                         & EA: $(\sigma_u)^{+1}$  & \num{-2.68} & \num{-1.56} &\num{-2.08} & \num{-0.55} & 0.16  & 0.34\\             
                                         & EA: satellite   & \num{-5.68} & \num{-5.36} & / & \num{-2.72} & \num{-2.52} & \num{-1.76}\\
     \hline\hline
    \end{tabular}
    \label{tab:orb-CASvsRHF}
\end{table}



\clearpage
\subsection{Additional results for \ce{O3}}
\subsubsection{Choice of active space}
The frontier RHF orbitals are summarized in Table \ref{fig:o3-act},
and the importance of different orbitals is assessed
in Table \ref{tab:o3-active-space}.

\begin{table}[!htb]
    \centering

    \renewcommand{\arraystretch}{1.5}
    \setlength{\tabcolsep}{.2cm}
    \caption{The nine RHF canonical orbitals formed by the $2p$ orbitals of oxygen for \ce{O3} obtained with the 6-31G basis set.}
    \begin{tabular}{cccc}
    \hline\hline
    Orbital & Figure & Orbital energy (Hartree) & Occupation \\\hline
\adjustbox{valign=c}{$5a_1$} & \adjustbox{valign=c}{\includegraphics[height=1.5cm]{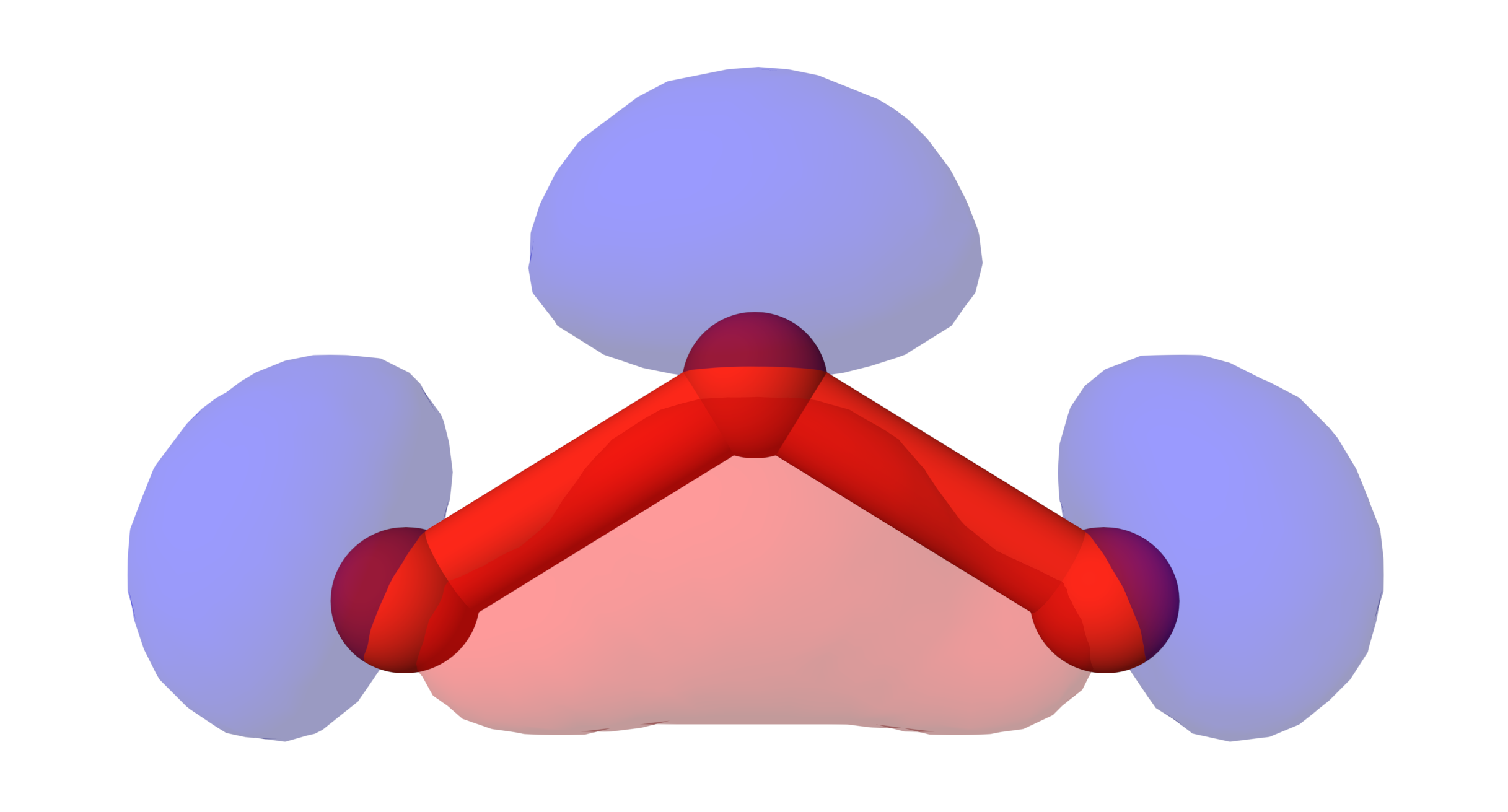}} &  \adjustbox{valign=c}{\num{-0.83815}} & 2\\\hline
\adjustbox{valign=c}{$3b_2$} & \adjustbox{valign=c}{\includegraphics[height=1.5cm]{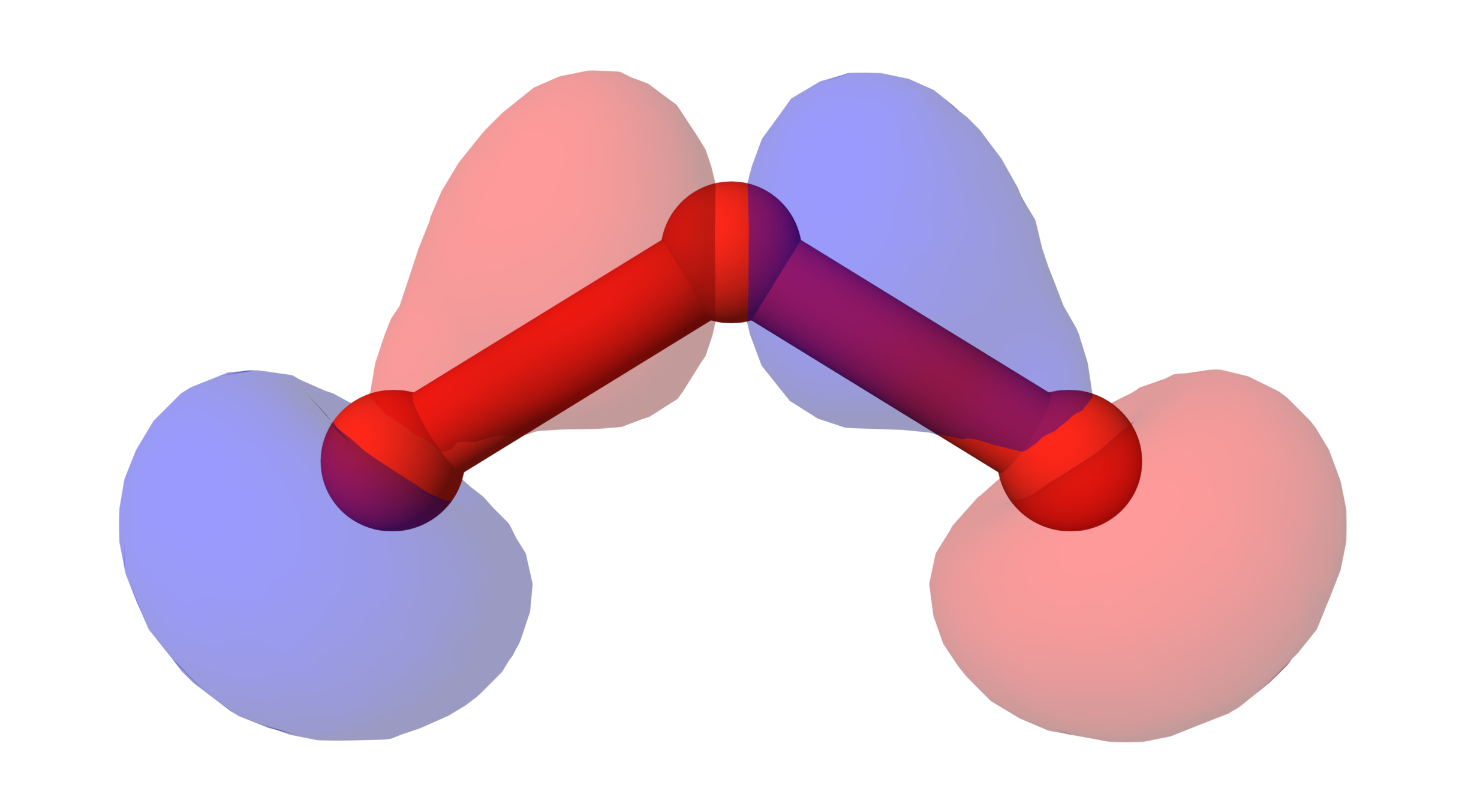}} &  \adjustbox{valign=c}{\num{-0.79362}} & 2\\\hline
\adjustbox{valign=c}{$1b_1$} & \adjustbox{valign=c}{\includegraphics[width=1.5cm, angle=90]{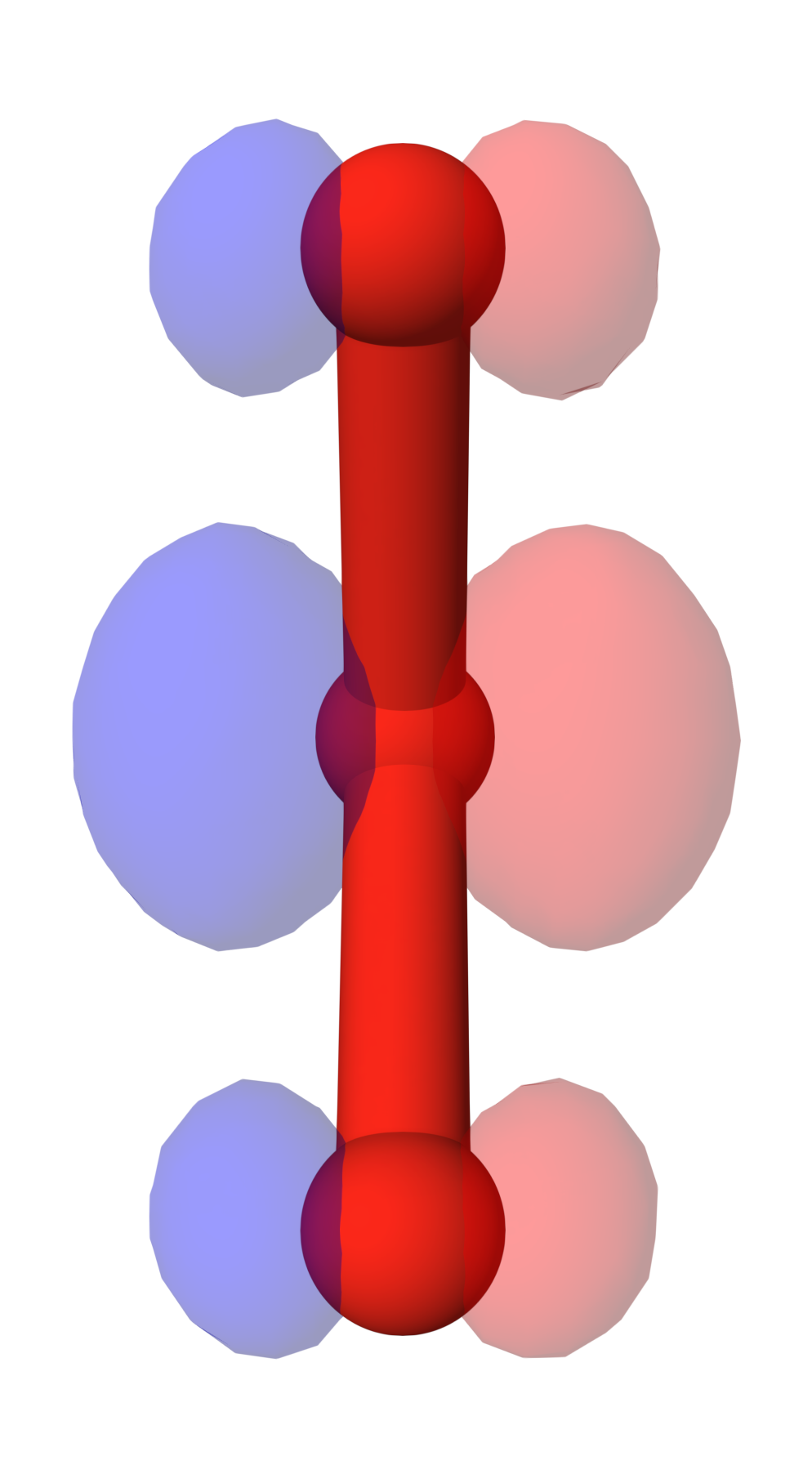}} &  \adjustbox{valign=c}{\num{-0.79183}} & 2\\\hline
\adjustbox{valign=c}{$4b_2$} & \adjustbox{valign=c}{\includegraphics[height=1.5cm]{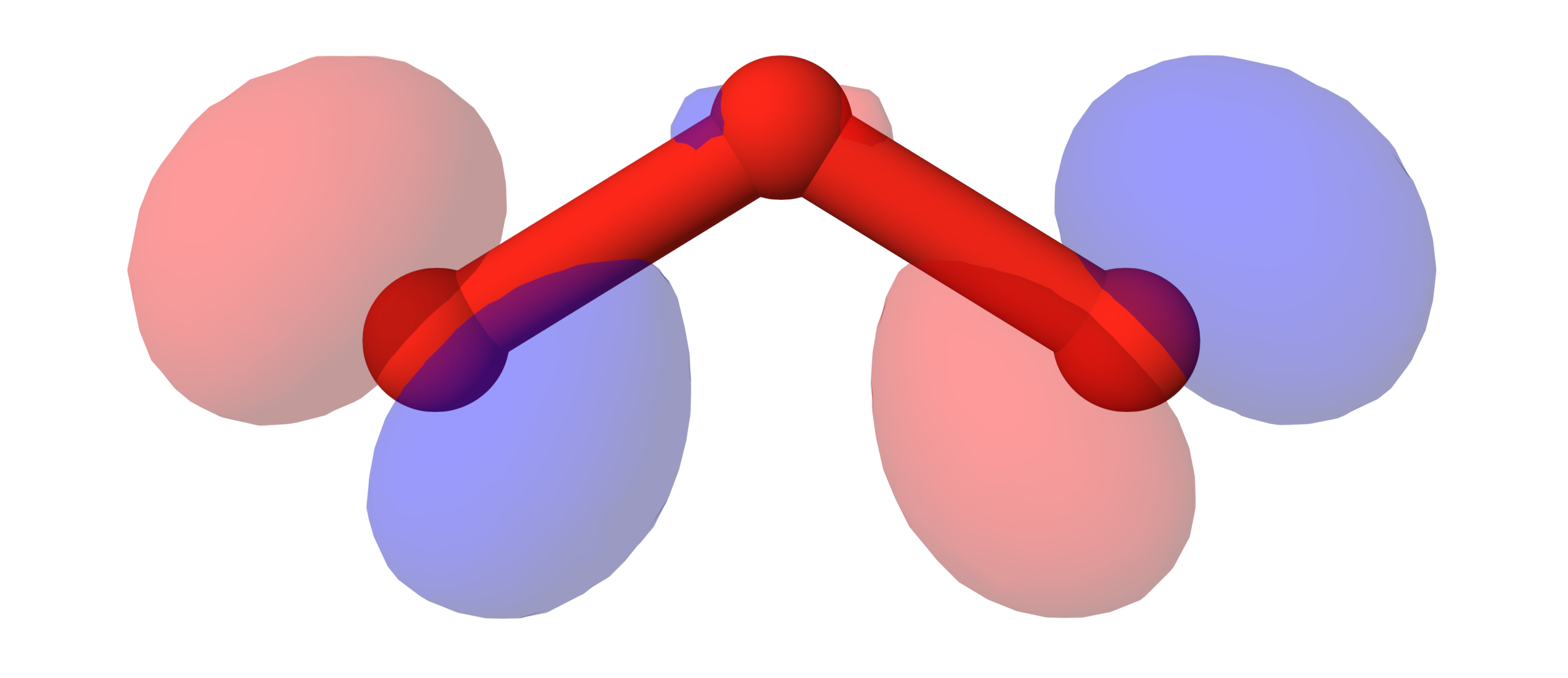}} &  \adjustbox{valign=c}{\num{-0.57902}} & 2\\\hline
\adjustbox{valign=c}{$6a_1$} & \adjustbox{valign=c}{\includegraphics[height=1.5cm]{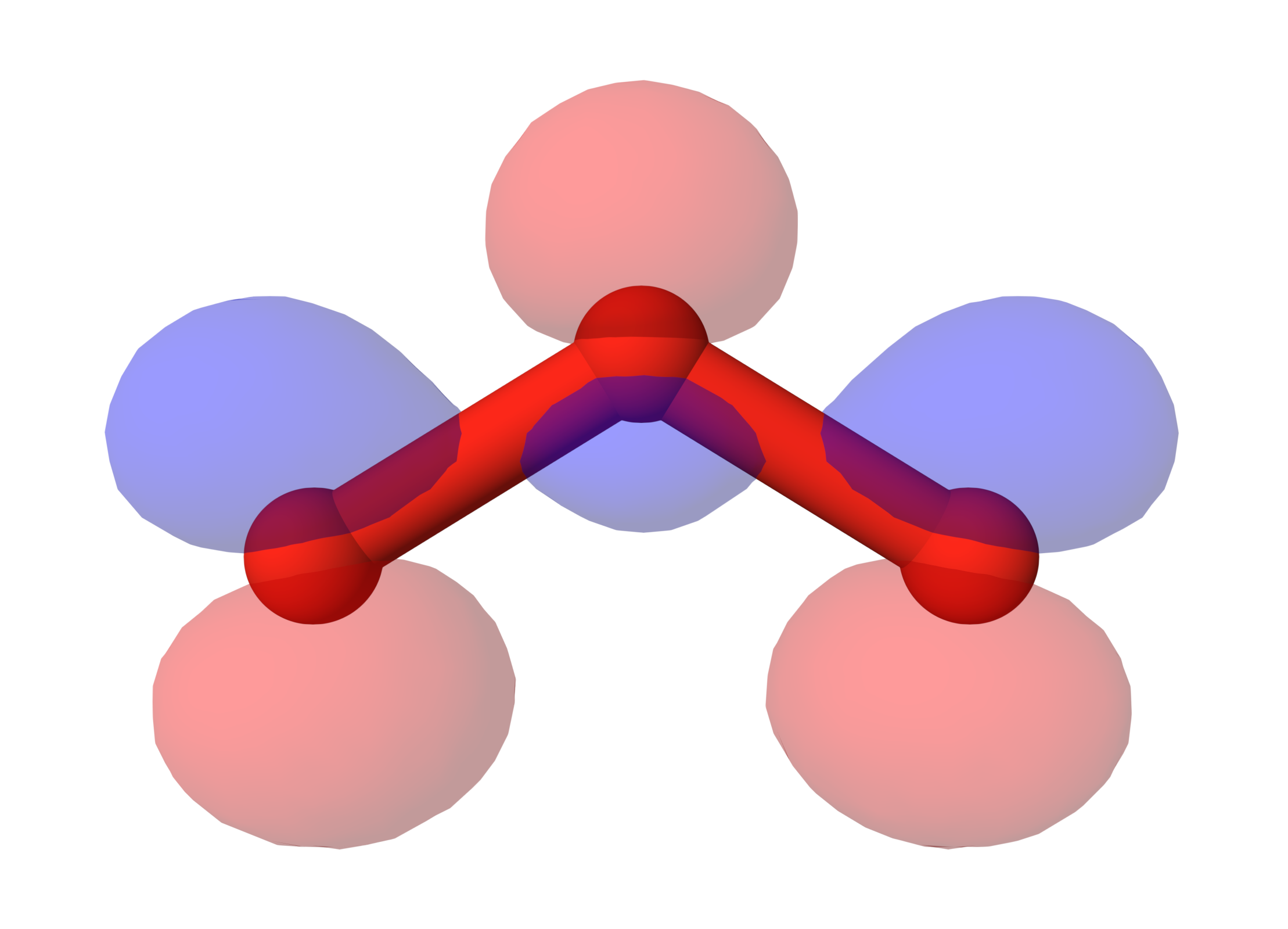}} &  \adjustbox{valign=c}{\num{-0.56193}} & 2\\\hline
\adjustbox{valign=c}{$1a_2$} & \adjustbox{valign=c}{\includegraphics[width=1.5cm, angle=90]{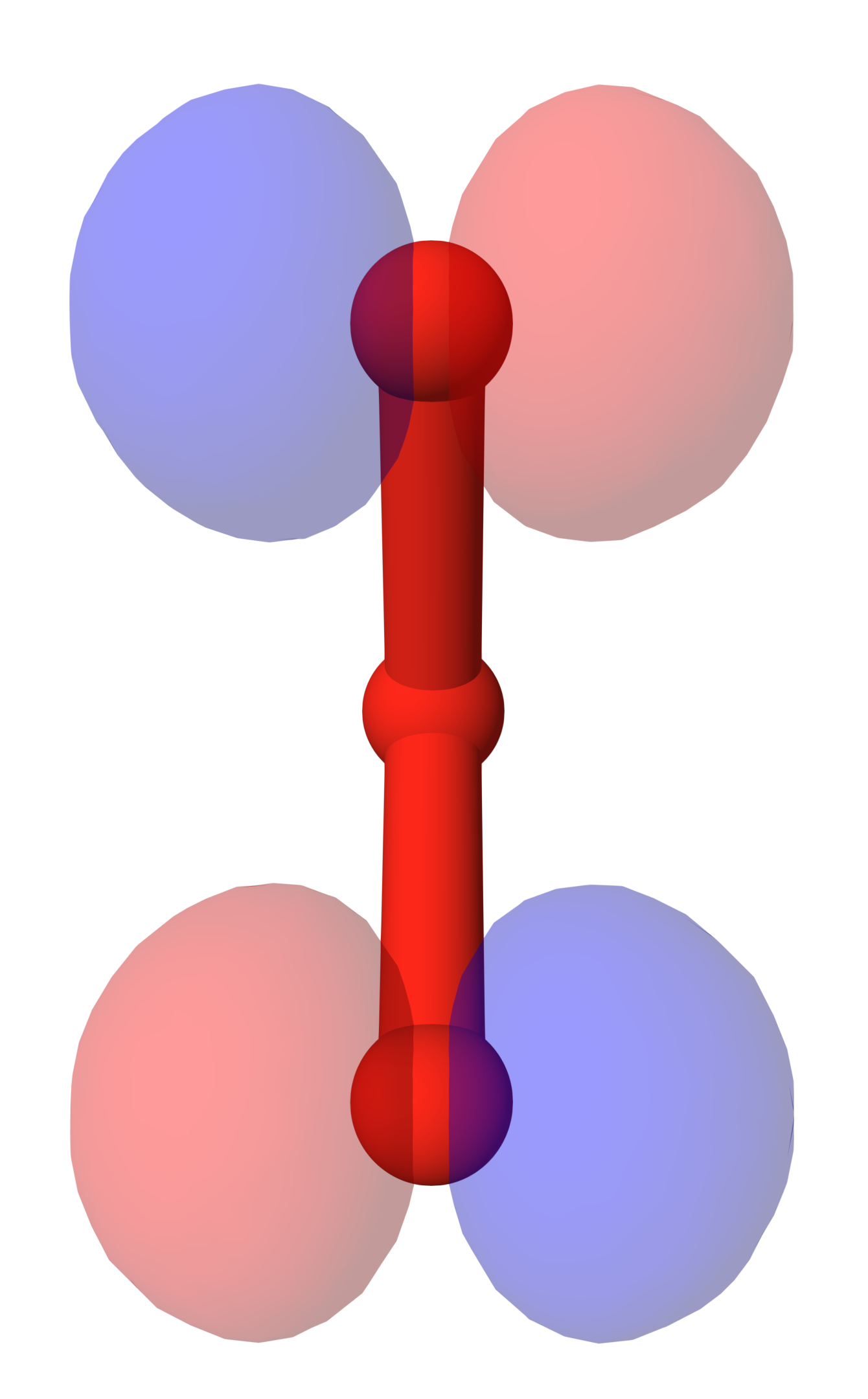}} &  \adjustbox{valign=c}{\num{-0.49558}} & 2\\\hline
\adjustbox{valign=c}{$2b_1$} & \adjustbox{valign=c}{\includegraphics[width=1.5cm, angle=90]{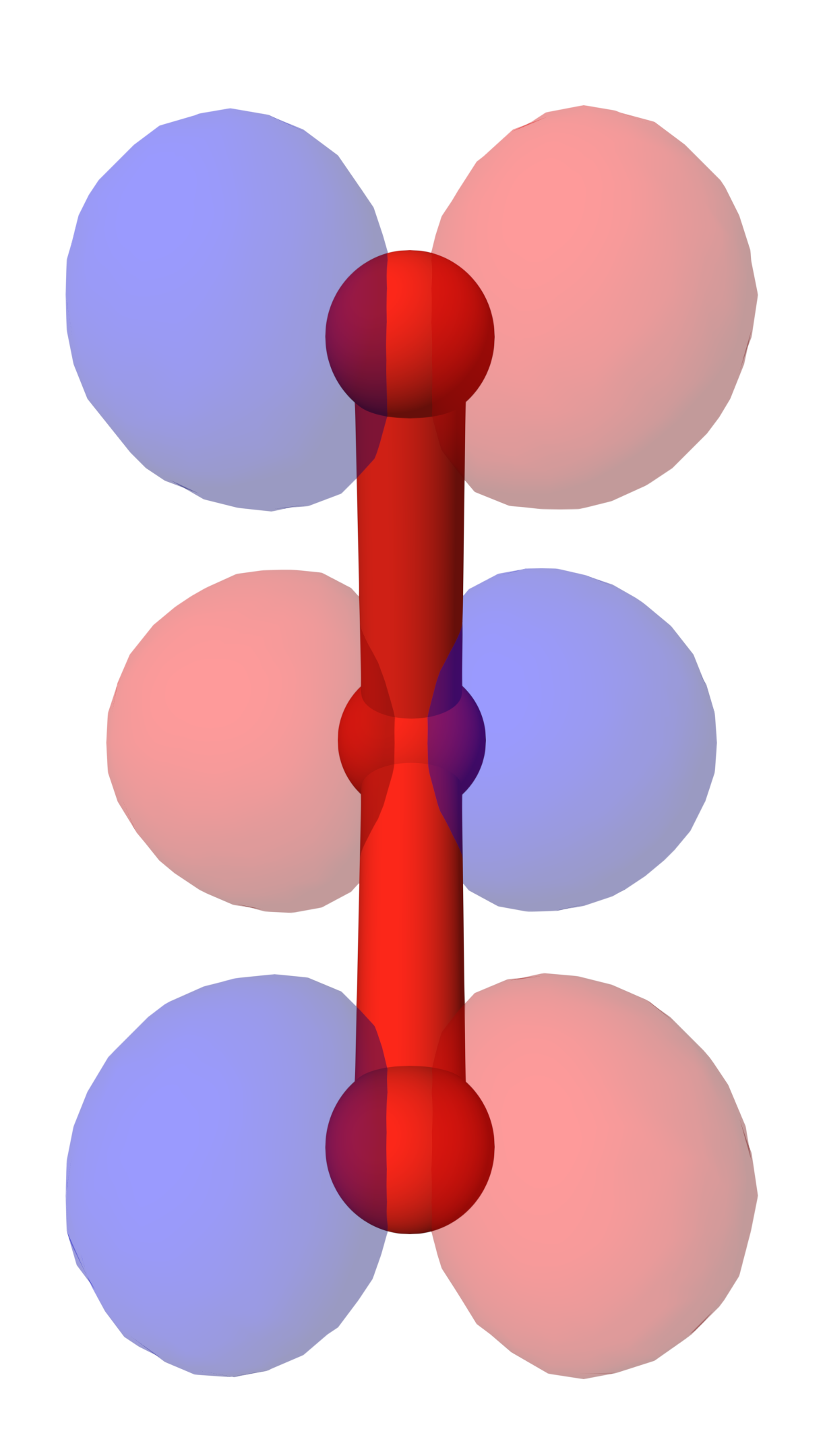}} &  \adjustbox{valign=c}{\num{-0.06794}} & 0\\\hline
\adjustbox{valign=c}{$7a_1$} & \adjustbox{valign=c}{\includegraphics[height=1.5cm]{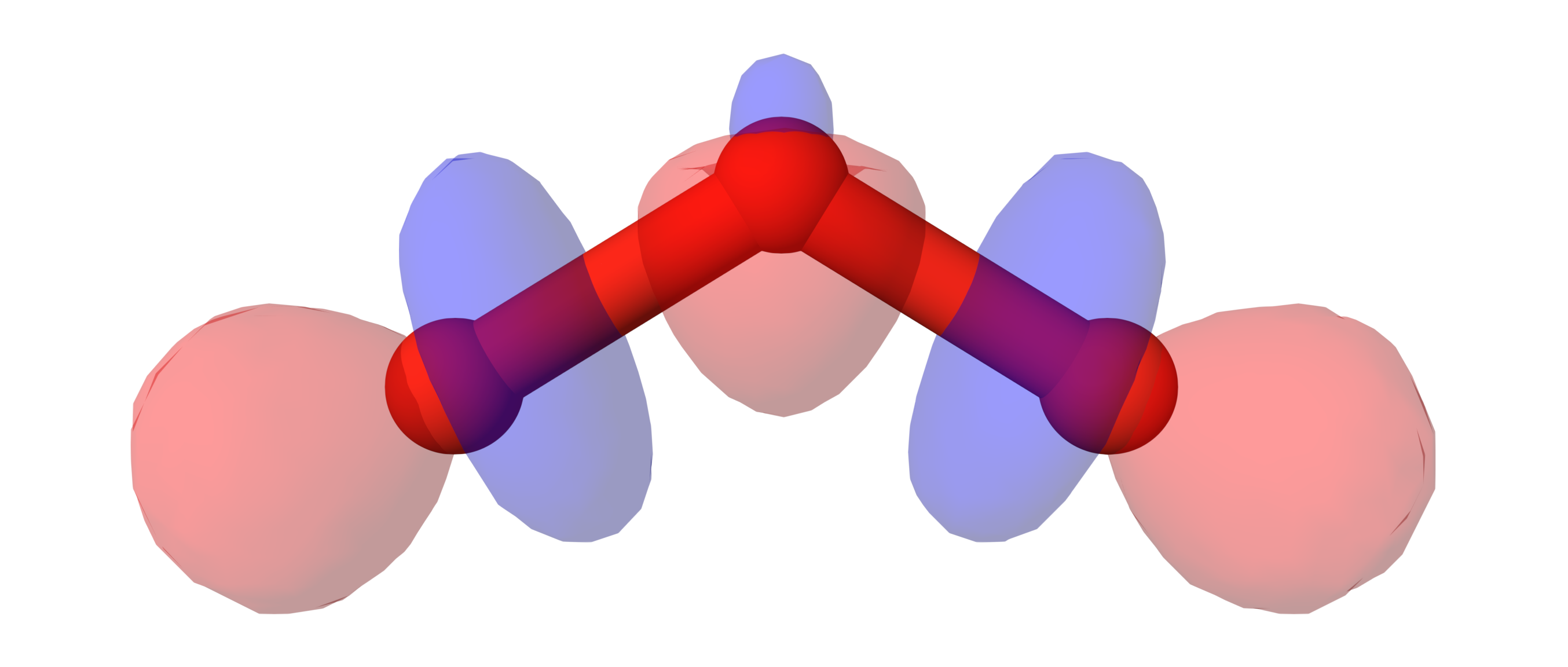}} &  \adjustbox{valign=c}{\num{0.26189}} & 0\\\hline
\adjustbox{valign=c}{$5b_2$} & \adjustbox{valign=c}{\includegraphics[height=1.5cm]{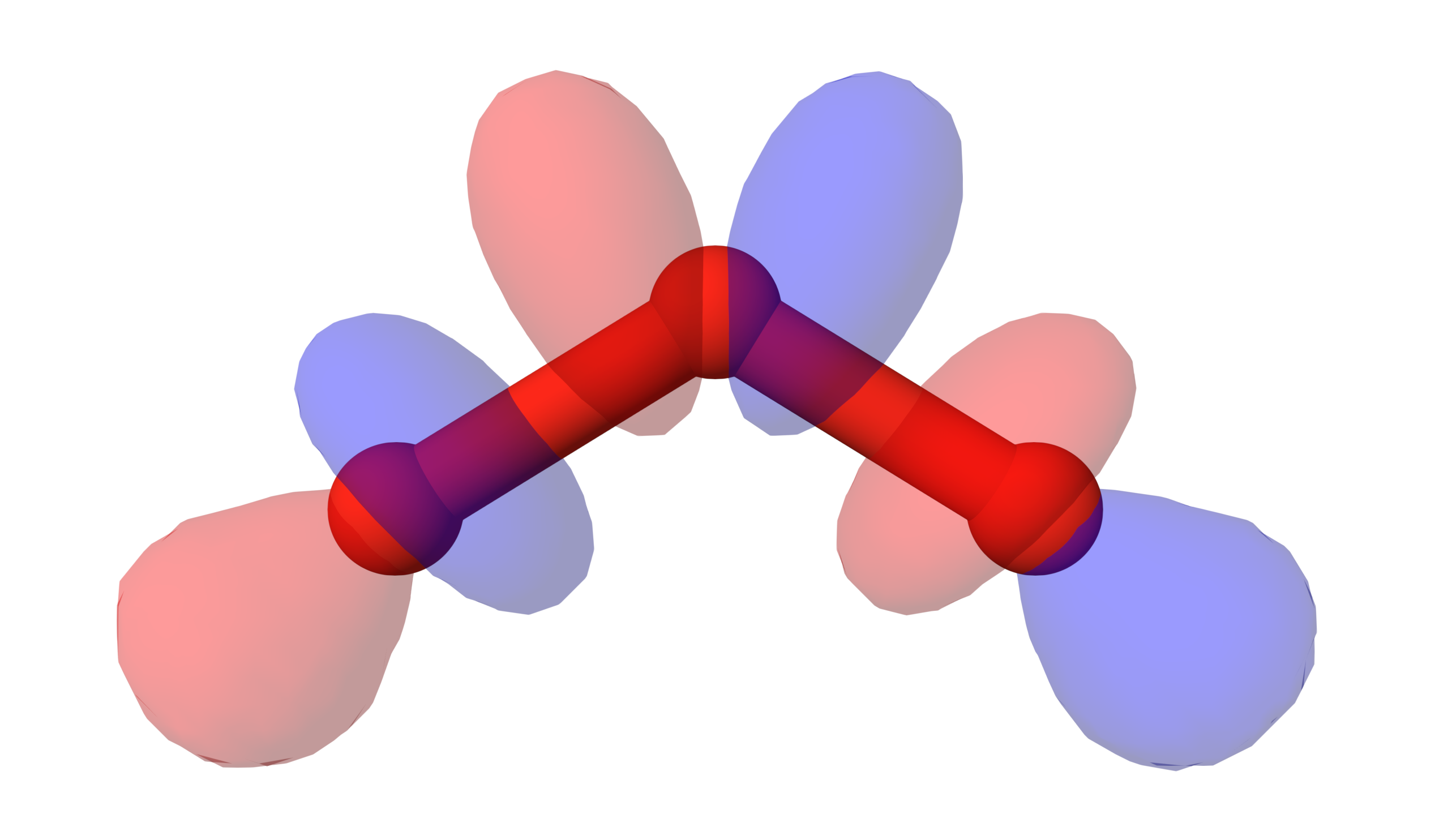}} &  \adjustbox{valign=c}{\num{0.36524}} & 0\\\hline\hline
    \end{tabular}

    \label{fig:o3-act}
\end{table}

\newpage

\begin{table}[!htb]
    \centering
    \renewcommand{\arraystretch}{1.3}
    \setlength{\tabcolsep}{4mm}
    \caption{Importance of different orbitals in computing the CASCI and MR-$GW$ vertical ionization energies (in eV) 
for the lowest three ionized states of \ce{O3}, assessed by
systematically enlarging the CAS(6,4) active space with
more orbitals. Calculations use the 6-31G basis set.}\label{tab:o3-active-space}
    \begin{tabular}{cc|ccc}
    \hline\hline
      Orbitals added & Method & $^2 A_1$ & $^2B_2$ & $^2 A_2$\\
     \hline
     \multirow{2}{*}{/}& CAS(6,4) & 14.38  & 14.67  & 15.12\\
     & MR-$GW$(6,4)  & 12.65 & 12.94 & 14.56 \\\hline
     \multirow{2}{*}{$5a_1$} & CAS(8,5)  & 14.08  & 14.22 & 15.12 \\
                             & MR-$GW$(8,5)      & 12.40  & 12.60 & 14.57\\\hline
    \multirow{2}{*}{$3b_2$}  & CAS(8,5)  & 14.10  & 14.49 & 15.11\\
                             & MR-$GW$(8,5)      & 12.45  & 12.73 & 14.54\\\hline
    \multirow{2}{*}{$1b_1$}  & CAS(8,5)  & 13.59  & 13.92 & 13.10\\
                             & MR-$GW$(8,5)      & 11.58  & 12.09 & 12.94\\\hline
    \multirow{2}{*}{$7a_1$}  & CAS(6,5)  & 14.38  & 14.59 & 15.02 \\
                             & MR-$GW$(6,5)      & 12.68  & 12.83 & 14.40  \\\hline
    \multirow{2}{*}{$5b_2$}  & CAS(6,5)  & 14.35  & 14.64 & 15.04 \\
                             & MR-$GW$(6,5)      & 12.64  & 12.95 & 14.41 \\\hline
    \multirow{2}{*}{$1b_1,7a_1$}    & CAS(8,6) & 13.78   &  14.10  &  13.28  \\
                                    & MR-$GW$(8,6)     & 11.93   &  12.33  &  13.09  \\\hline
    \multirow{2}{*}{$1b_1,5b_2$}   & CAS(8,6)  & 13.76  & 14.08 & 13.27 \\
                                   & MR-$GW$(8,6)      & 11.87   &  12.39  & 13.08\\\hline
    \multicolumn{2}{c|}{Exp.}      & 12.73     & 13.00  & 13.54\\
    \hline\hline
    \end{tabular}
\end{table}
\FloatBarrier

\clearpage
{\color{\mycolor}
\subsubsection{Vertical ionization energies at equilibrium}

Table~\ref{tab:o3-equilibrium-values} summarizes the vertical
ionization energies at the equilibrium $C_{2v}$ geometry,
$R(\ce{O-O})=1.278$\AA~ and $\theta(\ce{O-O-O})=116.8^\circ$, in the 6-31G basis set and provides
the numerical data underlying Fig.~\ref{fig:o3}.
The DMRG reproduces the experimental ordering $^2A_1 < {}^2B_2 < {}^2A_2$. The DMRG results are obtained with a maximum bond dimension
of $M=5000$, and the difference between the DMRG IPs with $M=3000$ and $M=5000$ is at most $0.02$ eV.
EOM-CCSD agrees with DMRG to within $0.08$~eV for all three states at this geometry.
%
%
For comparison, broken-symmetry $GW$@UHF at the same geometry
gives its three lowest ionization energies as 12.96, 14.18, and
14.31~eV. The underlying UHF solution breaks the $C_{2v}$
point-group symmetry, so these poles cannot be assigned
unambiguously to the $^2A_1$, $^2B_2$, and $^2A_2$ states in
the present calculations. 
}

\begin{table}[!htb]
    \centering
    \renewcommand{\arraystretch}{1.2}
    \setlength{\tabcolsep}{5mm}
    \caption{{\color{\mycolor}
    Vertical ionization energies (in eV)
    of \ce{O3} at its equilibrium geometry,
    $\theta(\ce{O-O-O})=116.8^\circ$. All calculated values use the
    6-31G basis set. 
    The CAS rows give the zeroth-order CASCI
    results represented by dashed lines in Fig.~\ref{fig:o3}, with active spaces defined
    there. DMRG uses a maximum bond dimension of $M=5000$.
    Additional EOM-CC2 and EOM-CCSD results, not plotted in
    Fig.~\ref{fig:o3}, are also listed for comparison.}
    }
    \label{tab:o3-equilibrium-values}
    \begin{tabular}{lccc}
    \hline\hline
    Method & $^2A_1$ & $^2B_2$ & $^2A_2$ \\
    \midrule
    KT              & 15.29 & 15.76 & 13.49 \\
    $GW$@RHF        & 13.25 & 13.52 & 13.20 \\
    Non-Dyson ADC(2)          & 10.69 & 10.79 & 13.05 \\
    Non-Dyson ADC(3)          & 12.61 & 12.79 & 12.49 \\
    CAS(6,4)        & 14.38 & 14.67 & 15.12 \\
    MR-$GW$(6,4)    & 12.65 & 12.94 & 14.56 \\
    CAS(8,5)        & 13.59 & 13.92 & 13.10 \\
    MR-$GW$(8,5)    & 11.58 & 12.09 & 12.94 \\
    CAS(8,6)        & 13.78 & 14.10 & 13.28 \\
    MR-$GW$(8,6)    & 11.93 & 12.33 & 13.09 \\
    EOM-CC2         & 10.54 & 10.61 & 13.20 \\
    EOM-CCSD        & 12.42 & 12.64 & 13.38 \\
    DMRG            & 12.34 & 12.59 & 13.34 \\
    Exp.\cite{wiesner_valence_2003} & 12.73 & 13.00 & 13.54 \\
    \hline\hline
    \end{tabular}
\end{table}
\FloatBarrier

\begin{figure}[!htb]
    \centering
    \includegraphics[width=0.98\linewidth]{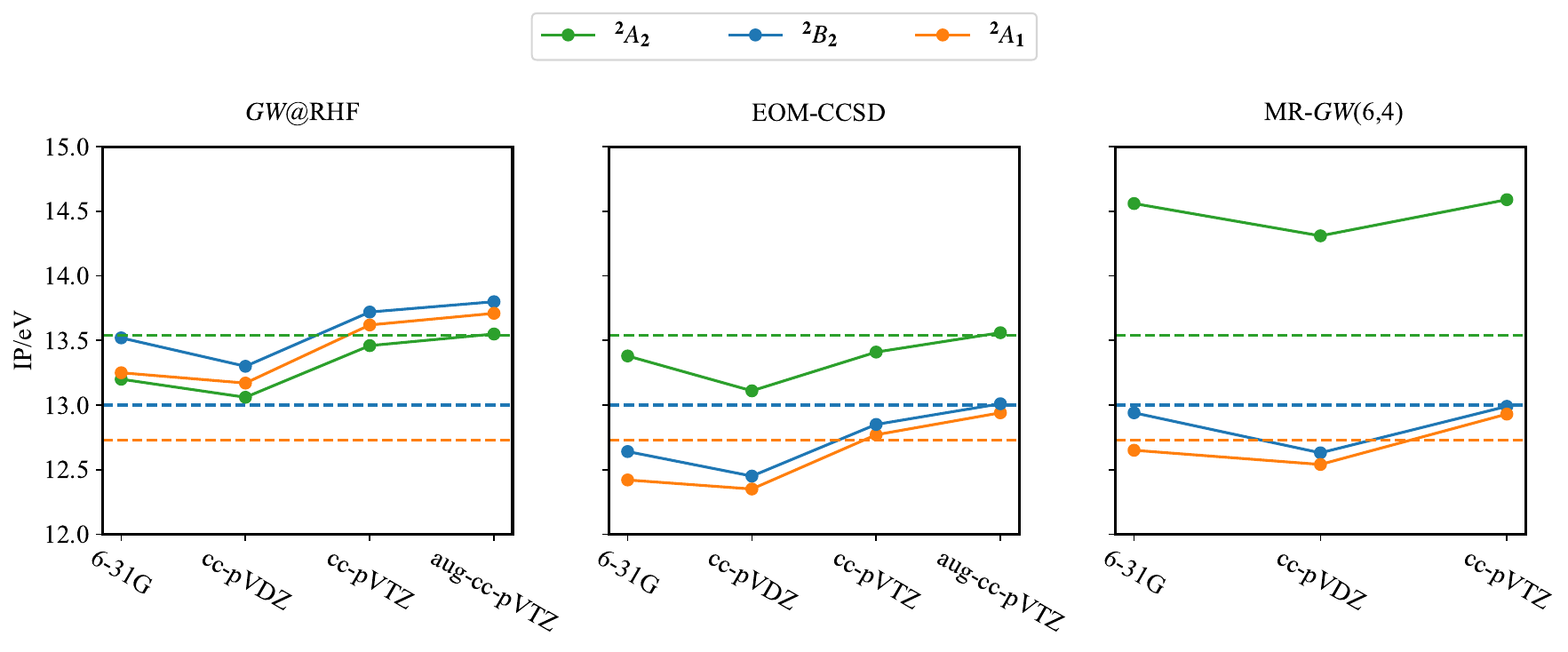}
    \caption{
	{\color{\mycolor}    
    Basis-set dependence of the lowest three vertical ionization energies of \ce{O3}. Panels (a)--(c) show $GW$@RHF, EOM-CCSD, and MR-$GW$(6,4), respectively. Solid curves are calculated values; color-matched dashed horizontal lines mark experiment. Data are taken from Table \ref{tab:o3-basis-values}.}
    }
    \label{fig:o3-basis-trends}
\end{figure}

\clearpage
{\color{\mycolor}
\subsubsection{Basis set dependence}
Basis-set effects on the vertical IPs of \ce{O3} leading to the
$^2A_1$, $^2B_2$, and $^2A_2$ states were investigated at the
$GW$@RHF, EOM-CCSD, and MR-$GW$(6,4) levels. The corresponding
energies are summarized in Table~\ref{tab:o3-basis-values} and
Fig.~\ref{fig:o3-basis-trends}.
Simultaneously enlarging the active space and the one-electron basis
is currently beyond the reach of our dense pilot implementation,
whose principal bottlenecks are the MR-RPA
diagonalizations. Such calculations will require lower-scaling
implementations. Nevertheless, all three methods exhibit similar basis-set dependence: 
the three IPs decrease from 6-31G to cc-pVDZ
and then increase as the basis is enlarged further, while their
relative ordering remains unchanged over the basis sets considered.
The consistency of this trend across $GW$@RHF, EOM-CCSD, and
MR-$GW$(6,4) suggests that it primarily reflects basis-set
incompleteness outside the active space. 
Although enlarging the active space will further shift the absolute IPs, 
MR-$GW$(6,4) therefore provides a useful estimate of the
qualitative basis-set dependence and state ordering.
}

\begin{table}[!hbt]
    \centering
    \renewcommand{\arraystretch}{1.3}
    \setlength{\tabcolsep}{4mm}
    \caption{
    {\color{\mycolor}
    Basis-set effects investigated at the $GW$@RHF, EOM-CCSD, and MR-$GW$(6,4) levels for IPs (in eV) to the $^2A_1$, $^2B_2$, and $^2A_2$ states of \ce{O3}.}
    }
    \begin{tabular}{c|c|ccc}
    \hline\hline
                               &  Basis set   &  $^2A_1$ & $^2B_2$ & $^2A_2$ \\
     \hline
     \multirow{4}{*}{$GW$@RHF}  & 6-31G       &  13.25  & 13.52  & 13.20\\
                                & cc-pVDZ   & 13.17 & 13.30 & 13.06 \\
                                & cc-pVTZ   & 13.62 & 13.72 & 13.46 \\
                                & aug-cc-pVTZ & 13.71 & 13.80 & 13.55 \\
     \hline
     \multirow{4}{*}{EOM-CCSD}  & 6-31G & 12.42 & 12.64 & 13.38 \\
                                & cc-pVDZ & 12.35 & 12.45 & 13.11 \\
                                & cc-pVTZ & 12.77 & 12.85 & 13.41 \\
                                & aug-cc-pVTZ & 12.94 & 13.01 & 13.56 \\
    \hline
    \multirow{3}{*}{MR-$GW$(6,4)}   & 6-31G       & 12.65 & 12.94 & 14.56   \\
                               & cc-pVDZ     & 12.54  & 12.63  & 14.31 \\
                               & cc-pVTZ     & 12.93  & 12.99   & 14.59 \\
    \hline
    \multicolumn{2}{c|}{Exp.}   & 12.73     & 13.00  & 13.54 \\
    \hline\hline
    \end{tabular}
    \label{tab:o3-basis-values}
\end{table}

\FloatBarrier

\begin{figure}[!htb]
    \centering
    \includegraphics[width=0.9\linewidth]{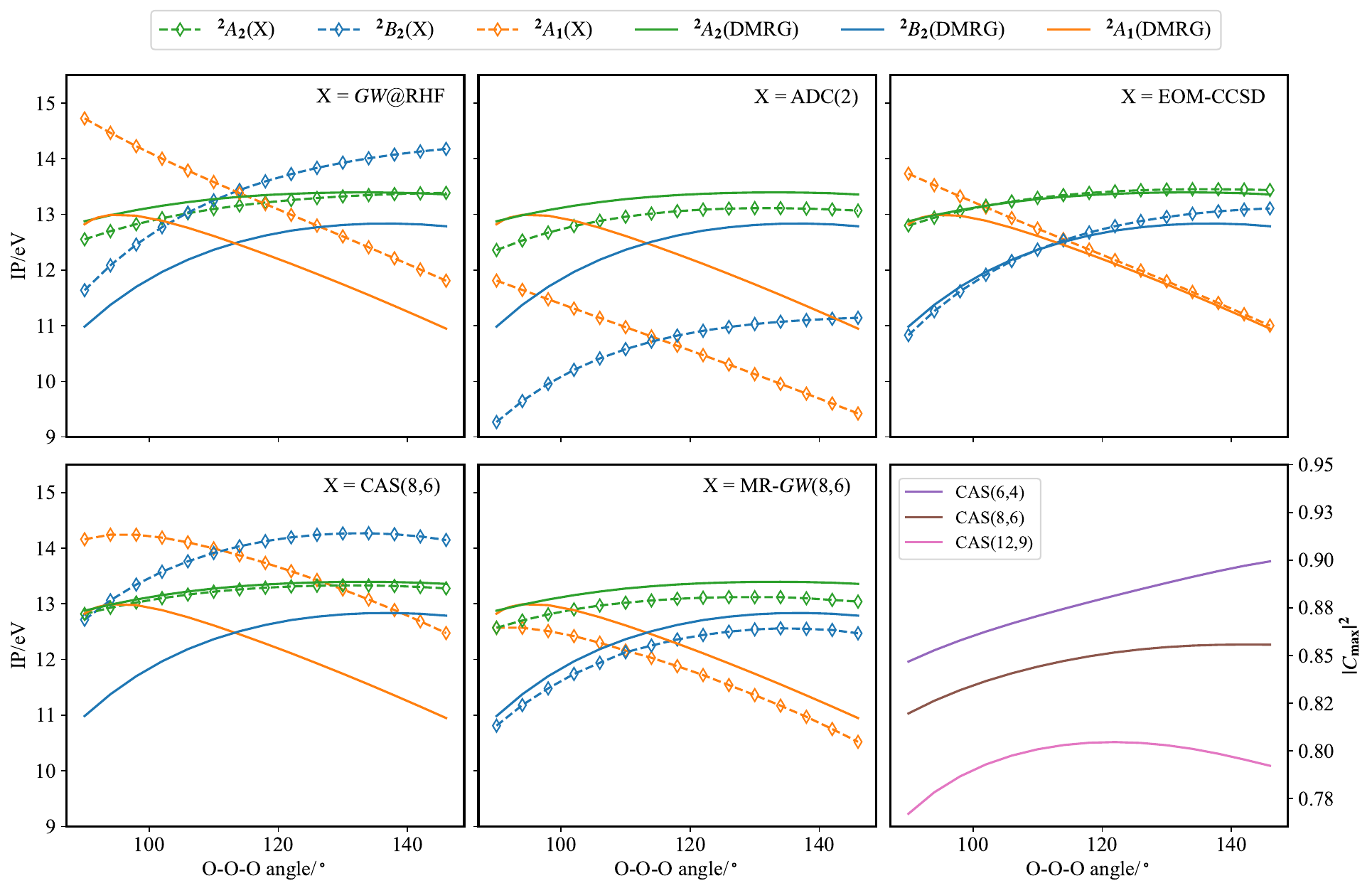}
    \caption{{\color{\mycolor}
    Vertical ionization energies (IP, in eV) and reference-state
    multiconfigurational character of \ce{O3} along the \ce{O-O-O}
    bending coordinate ($90^\circ$--$146^\circ$) in the 6-31G basis set.
    Both \ce{O-O} bond lengths are held fixed at their equilibrium
    values throughout the scan.
    In the five IP panels, dashed lines with open diamonds denote the
    method $X$ indicated in each panel, while solid lines give the
    DMRG references with maximum bond dimension $M=5000$.
    Orange, blue, and green identify the $^2A_1$, $^2B_2$, and $^2A_2$
    ionized states, respectively. The lower-right panel shows the largest
    squared CI coefficient, $|C_{\max}|^2=\max_I|C_I|^2$, of the neutral
    CAS wave function in three active spaces. CAS(6,4) and CAS(8,6)
    are defined in Fig.~\ref{fig:o3}; CAS(12,9) contains the nine
    orbitals listed in Table~\ref{fig:o3-act}.
	Data are available in Ref. \cite{datarepo}.    
    }}
    \label{fig:o3-bending-scan}
\end{figure}

{\color{\mycolor}
\subsubsection{Ionization-energy curves along the \ce{O-O-O} bending coordinate}

At fixed \ce{O-O} bond lengths, Fig.~\ref{fig:o3-bending-scan} connects
the angle dependence of the ionization energies to the multiconfigurational
character of the neutral reference. In particular, $|C_{\max}|^2$ decreases on
bending from equilibrium toward $90^\circ$, 
indicating a growing departure from a single-determinant description in the chosen orbital
representation. The DMRG curves show pronounced, state-dependent
variations of the $^2A_1$ and $^2B_2$ IPs, whereas the $^2A_2$ IP
varies more weakly.
$GW$@RHF substantially overestimates the $^2A_1$ and $^2B_2$ IPs, while
ADC(2) strongly underestimates both IPs. EOM-CCSD closely follows
DMRG near and beyond the equilibrium angle, but its $^2A_1$ error
increases to $0.91$~eV at $90^\circ$.
The CAS(8,6) reference already captures the main angular trends,
but its $^2A_1$ and $^2B_2$ IPs are substantially too high.
Screening in MR-$GW$(8,6) improves their absolute energies while
retaining the overall angular dependence. At $90^\circ$, the
$^2A_1$ error is reduced to -$0.25$~eV, compared with $1.90$~eV for
$GW$@RHF. 
The more faithful angular trends of
MR-$GW$ illustrate how the interacting reference and dynamical screening work together across the bending coordinate.
}

\end{document}